\documentclass[twocolumn,superscriptaddress,secnumarabic,amssymb,nobibnotes,aps,bm,prb,floatfix,longbibliography]{revtex4-2}
\usepackage[english]{babel}
\usepackage[latin9]{inputenc}
\usepackage{graphicx}
\usepackage{multirow}
\usepackage[usenames,dvipsnames]{xcolor}
\usepackage{bm}
\usepackage{mathtools}
\usepackage{etoolbox}
\usepackage{eqnarray}
\usepackage{amsmath}
\usepackage{rotating}
\usepackage{gensymb}
\definecolor{oceanboatblue}{rgb}{0.0, 0.47, 0.75}
\definecolor{orange}{rgb}{1,0.5,0}
\definecolor{goodgreen}{rgb}{0.1,0.5,0}
\definecolor{goodred}{rgb}{0.7,0,0}
\usepackage[colorlinks,urlcolor=goodgreen,citecolor=blue,linkcolor=goodred]{hyperref}
\usepackage{cleveref}
\usepackage{lineno}
\setpagewiselinenumbers
\modulolinenumbers[5]

\usepackage{tikz}
\usepackage{tocvsec2}
\usepackage{scrextend}
\makeatletter

\newcommand{\bea}{\begin{eqnarray}}
\newcommand{\bse}{\begin{subequations}}
\newcommand{\ese}{\end{subequations}}
\newcommand{\eea}{\end{eqnarray}}
\newcommand{\beq}{\begin{equation}}
\newcommand{\eeq}{\end{equation}}
\newcommand{\ci}{\mathrm{i}}
\newcommand{\ketLR}[1]{\left| #1 \right\rangle}

\newcommand{\ee}{\text{e}}
\renewcommand{\exp}[1]{\ee^{#1}}
\newcommand{\comment}[1]{}
\newcommand{\ket}[1]{| #1 \rangle}

\newcommand{\bra}[1]{\langle #1 |}

\usepackage[normalem]{ulem}

\makeatother

\begin{document}
\title{Parallel spin transport and holonomy in non-Euclidean curved circuits on a spherical two-dimensional electron gas}

\author{E. J. Rodr\'{\i}guez}
\email{erfernandez@us.es}
\affiliation{Departamento de F\'isica Aplicada II, Universidad de Sevilla, E-41012 Sevilla, Spain}

\author{A. A. Reynoso}
\email{reynoso@cab.cnea.gov.ar}
\affiliation{Instituto Balseiro and Centro At\'omico Bariloche, Comisi\'on Nacional de Energ\'ia At\'omica, 8400 Bariloche, Argentina}

\author{J. P. Baltan\'as}
\email{baltanas@us.es}
\affiliation{Departamento de F\'isica Aplicada II, Universidad de Sevilla, E-41012 Sevilla, Spain}

\author{D. Bercioux}
\email{dario.bercioux@dipc.org}
\affiliation{Donostia International Physics Center (DIPC), 20018 Donostia--San Sebasti\'an, Spain}
\affiliation{IKERBASQUE, Basque Foundation for Science, Plaza Euskadi 5
48009 Bilbao, Spain}

\author{D. Frustaglia}
\email{frustaglia@us.es}
\affiliation{Departamento de F\'isica Aplicada II, Universidad de Sevilla, E-41012 Sevilla, Spain}

\begin{abstract}

The quantum conductance of one-dimensional (1D) circuits built on flat (Euclidean) two-dimensional electron gases (2DEGs) is known to display a symmetric response to the inversion of Rashba spin-orbit coupling fields in Aharonov-Casher (AC) interference patterns. Here, we show that this symmetry breaks down in curved (non-Euclidean) 1D circuits defined on spherical 2DEGs. We demonstrate that this is a consequence of parallel transport and holonomy of the electronic spin on the surface of the sphere, and that a symmetric response can be recovered when considering the parallel transport condition as an offset shifting the AC pattern. We discuss 1D triangular circuits defined along geodesic arcs on the sphere as a case study, and generalize it to regular polygons and parallel curves of given latitude.

\end{abstract}
\date{\today}
\maketitle
\section{Introduction}
Rashba spin-orbit interaction ~\cite{Rashba1960,Bychkov1984} is a mechanism that couples the electronic orbital and spin degrees of freedom, naturally arising in two-dimensional electron gases (2DEGs) with structural inversion asymmetry and widely known as an effective tool for the manipulation and control of electronic spins~\cite{Nitta1997,Manchon2015}. Particular attention has been paid to geometric phases that develop when spin carriers move in materials where Rashba spin-orbit coupling (SOC) occurs. The effects of such quantum phases have been experimentally observed in the transport properties of loop-like interferometers, customarily implemented as one-dimensional (1D) circuits lying in Euclidean (flat) 2DEGs~\cite{Bercioux_2015,Frustaglia2020}. Moreover, recent experimental progress in curved electronic, magnetic and atomic low-dimensional platforms~\cite{Gentile2022,Streubel2016,Tononi2023} has triggered the study of novel phenomena at the nanoscale anticipated by early works on quantum dynamics in curved spaces, as two-dimensional (2D) manifolds embedded in three-dimensional (3D) space~\cite{Jensen1971,daCosta1982}. Several efforts are presently devoted to the study of geometric quantum potentials and topological states in non-Euclidean 2DEGs with positive/negative Gaussian curvatures corresponding to elliptic/hyperbolic geometries, respectively~\cite{Ortix2010, Rosdahl2015, Kozlovsky2020, Furst2024, Grass_2024}. Among other findings, the absence of weak-localization corrections to the conductance has been predicted in hyperbolic 2DEGs due to a statistical deficit of correlated time-reversed paths~\cite{Curtis2023}.

Besides, significant progress has been made over the last decade in the study of spin carrier dynamics in curved spaces, especially following the work by Ortix~\cite{ortix2015} which paved the way for subsequent studies, including this one. Ortix~\cite{ortix2015} derived a Hamiltonian for non-relativistic spin carriers that propagate along 1D circuits of arbitrary shape embedded in 3D space subject to SOC originated by an electric field. This approach reproduced known results for 1D circuits of constant curvature, such as Rashba rings~\cite{Meijer2002, Frustaglia2004} built on flat (i.e., Euclidean) 2DEGs, and it can also be applied to other flat circuits with changing curvature, such as ellipses~\cite{Ying2016} and polygons~\cite{Rodriguez2021}.
It has been shown that the quantum conductance of Euclidean 1D circuits exhibits a symmetric response under the inversion of the Rashba SOC's sign in Aharonov-Casher (AC) spin interference patterns~\cite{AharonovCasher1984, MeirGefenEntin-Wohlman89, MathurStone92}. Here, we show that the symmetric response of the quantum conductance under Rashba SOC inversion does not hold in non-Euclidean 1D circuits build on curved 2DEGs.

To this aim, we examine curved 1D Rashba circuits built on a spherical 2DEG~\cite{Chang2013}. As a result, we find that the dynamics of the spin carriers and the corresponding AC phases acquired during propagation in these circuits differ significantly when the sign of the Rashba SOC is inverted. This leads to an asymmetric response in the quantum conductance. We demonstrate that this is a consequence of parallel spin transport and holonomy~\cite{doCarmo2016,Dandoloff1989,Urbantke1991,Dandoloff1992} on the curved surface.
We find that the Rashba SOC produced by a radial electric field on a spherical 2DEG (originated by an asymmetric radial confining potential~\cite{Chang2013} or surface strain~\cite{Salamone2022}, formally equivalent to a central electric charge) serves as a geometric connection for spin. This connection facilitates the realization of parallel spin transport for a particular setting, with SOC strength directly proportional to the sphere's curvature. Notice that this condition trivially reduces to a vanishing Rashba SOC in flat 2DEGs. A symmetric response of AC phases and conductance to Rashba SOC in 1D circuits on spherical 2DEGs is reestablished when taking the parallel transport condition as an offset that shifts the AC pattern. To our knowledge, these outcomes passed unnoticed in previous works on spin-carrier dynamics in curved spaces~\cite{ortix2015, Chang2013, Liang2018, Liang2020,Cheng2011,Chen2013}.

The article is organized as follows. We start by introducing geodesic 1D quantum wires on a Rashba sphere in Sec.~\ref{sec:Geo1Dwire}. Based on this, we define parallel spin transport and demonstrate its physical realization by means of Rashba SOC in Sec.~\ref{sec:PST}. In Sec.~\ref{sec:TriCircs}, we discuss non-adiabatic spin carrier dynamics in regular triangular circuits of different sizes defined along geodesic curves as a case study, showing the response of AC phases and quantum conductance to curvature. In the Appendices, we generalize our approach to regular polygonal circuits and, eventually, to parallel curves corresponding to different latitudes on a Rashba sphere.

\section{Geodesic 1D quantum wires on a Rashba sphere}
\label{sec:Geo1Dwire}
 
\begin{figure}[!h]
\centering
    \includegraphics[width=0.6\columnwidth]{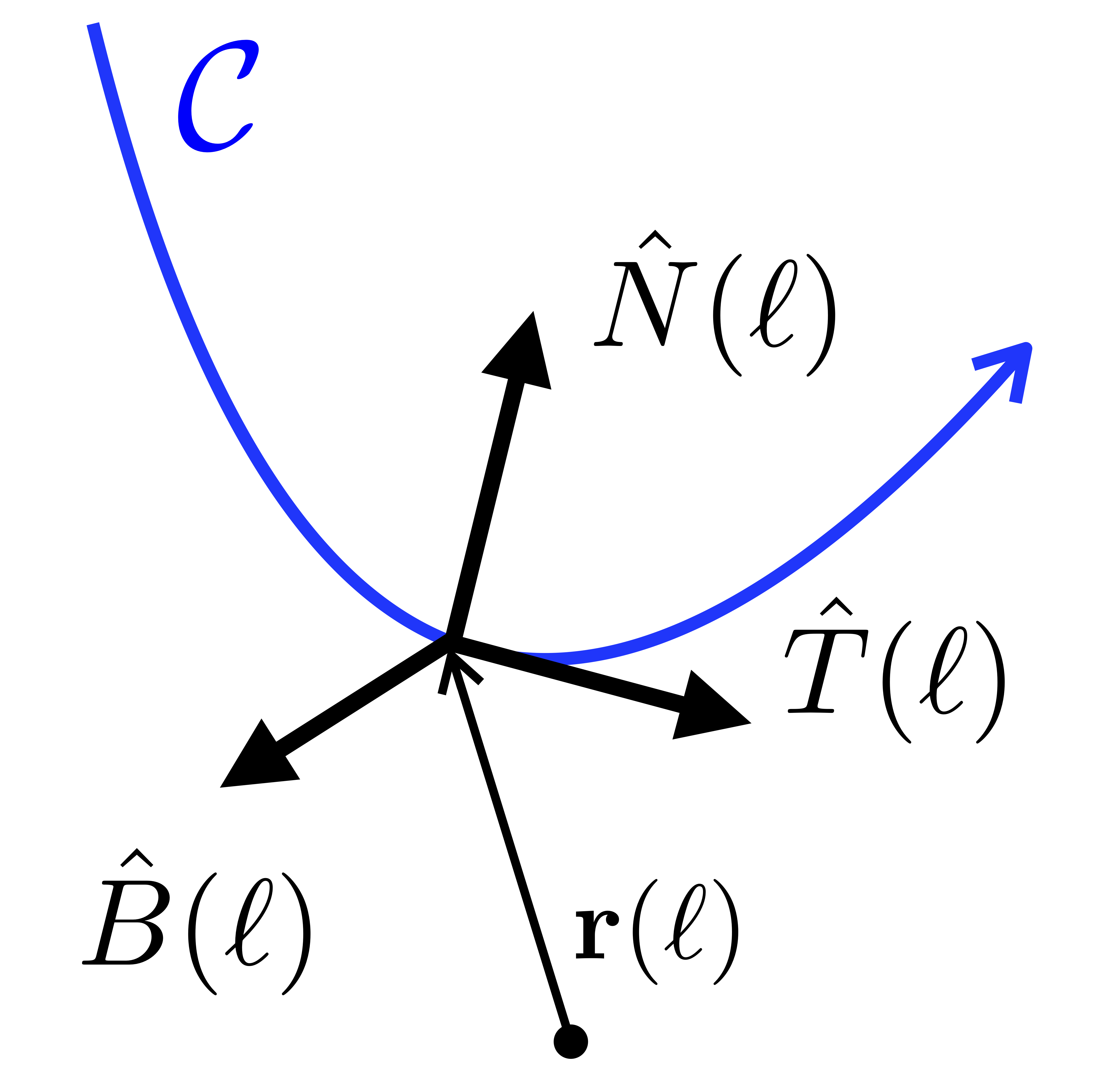}
    \caption{\label{TNB} 1D curve $\mathcal{C}$ embedded in 3D space and parametrized by $\mathbf{r}(\ell)$, with $\ell$ the arclength. It displays the local Frenet-Serret triad $\{\hat{T}(\ell),\hat{N}(\ell),\hat{B}(\ell)\}$.} 
\end{figure}

We model spin carrier dynamics along curved 1D quantum circuits by following the works of Jensen and Koppe~\cite{Jensen1971}, da~Costa~\cite{daCosta1982}, and Ortix~\cite{ortix2015}. 
The Hamiltonian for non-relativistic electrons in a flat 2DEG subject to SOC produced by a uniform electric field reads
\begin{equation} \label{H0}
    H=\frac{{\bf p}^2}{2m}+\frac{1}{\hbar} \boldsymbol{\alpha}\cdot(\boldsymbol{\sigma}\times{\bf p}),
\end{equation}
with ${\bf p}=-i\hbar \boldsymbol{\nabla}$ the in-plane momentum operator, $\boldsymbol{\sigma}$ the vector of Pauli matrices, and $\boldsymbol{\alpha}$ a vector proportional to the electric field that determines the SOC strength and axis. For $\boldsymbol{\alpha}$ perpendicular to the 2DEG, the second term of Eq.~\eqref{H0} corresponds to the standard Rashba SOC Hamiltonian~\cite{Bercioux_2015}. Equation~\eqref{H0} can be generalized to curved 2D surfaces by introducing appropriate metric tensors and geometric connections~\cite{ortix2015}.
We are interested in 1D circuits defined by quantum wires arranged along a curve $\mathcal{C}$ parametrized by ${\bf r}(\ell)$ with $\ell$ the arclength. To this aim, we define the right-handed triad of unit vectors tangent, normal, and binormal to $\mathcal{C}$, $\{\hat{T}(\ell)=\partial_\ell{\bf r}(\ell),\hat{N}(\ell),\hat{B}(\ell)\}$ as shown in Fig.~\ref{TNB}, obeying the Frenet-Serret equations
\begin{subequations}
\begin{align}
\partial_\ell\hat{T}(\ell)=&\kappa(\ell)\hat{N}(\ell),\\
\partial_\ell\hat{N}(\ell)=&-\kappa(\ell)\hat{T}(\ell)+\tau(\ell)\hat{B}(\ell),\\
\partial_\ell\hat{B}(\ell)=&-\tau(\ell)\hat{N}(\ell),
\end{align}
\end{subequations}
with $\kappa(\ell)$ and $\tau(\ell)$ the local curvature and torsion of $\mathcal{C}$, respectively. We now define $\sigma_{T,N,B}$ and $\alpha_{T,N,B}$ as the
projections of $\boldsymbol{\sigma}$ and $\boldsymbol{\alpha}$ along
the local triad, which in general depend on $\ell$. For the particular
case of constant $\alpha_{N}$ and $\alpha_{B}$, one finds that
electrons propagating along planar curves $\mathcal{C}_0$ (i.e., with
vanishing torsion) respond to the Hamiltonian~\cite{ortix2015}
\begin{align}
    H_{\mathcal{C}_0}=&-\frac{\hbar^2}{2m}\left( \partial_\ell^2+\frac{\kappa^2(\ell)}{4} \right)
    +i\alpha_B\left( \sigma_N\partial_\ell-\sigma_T\frac{\kappa(\ell)}{2}\right) \nonumber \\ \label{H1}
    &-i\alpha_N \sigma_B\partial_\ell,
\end{align}
where we assume that $\sigma_{T,N,B}$ implicitly depends on $\ell$ to simplify the notation. The SOC terms in Eq.~\eqref{H1} define an effective, momentum-dependent magnetic field (or magnetic texture) interacting with the spin via a Zeeman-type coupling that respects time-reversal symmetry.
Moreover, note that $\alpha_{T}(\ell)$ does not lead to any SOC term in Eq.~\eqref{H1} since this component is parallel to the momentum. 
Formally, the derivation of the 1D Hamiltonian~\eqref{H1} assumes the existence of higher dimensions restricted by a thin-wall quantization procedure that takes into account strong confining potentials in the normal~($\hat{N}$) and binormal~($\hat{B}$) directions~\cite{ortix2015}.
For 1D Rashba rings of radius $r$ on flat 2DEGs (with vanishing $\alpha_N$, constant curvature $\kappa=1/r$, and radial effective magnetic texture along $\hat{N}$), Eq.~\eqref{H1} reduces to the correct Hamiltonian discussed in the literature~\cite{Meijer2002, Frustaglia2004}. Moreover, Eq.~\eqref{H1} also incorporates a scalar quantum potential of geometric origin (proportional to the square of the local curvature) in its kinetic term~\cite{Jensen1971}. This contribution is usually disregarded in 1D Rashba rings since it reduces to a constant energy offset. Its effects are also minimized in the semiclassical limit, typically valid on the mesoscopic scale~\cite{Ying2020}. Still, the scalar geometric quantum potential can lead to interesting phenomena in curved materials at the nanoscale, as discussed in Ref.~\cite{Gentile2022}. 

\begin{figure}[!h]
\centering
    \includegraphics[width=0.6\columnwidth]{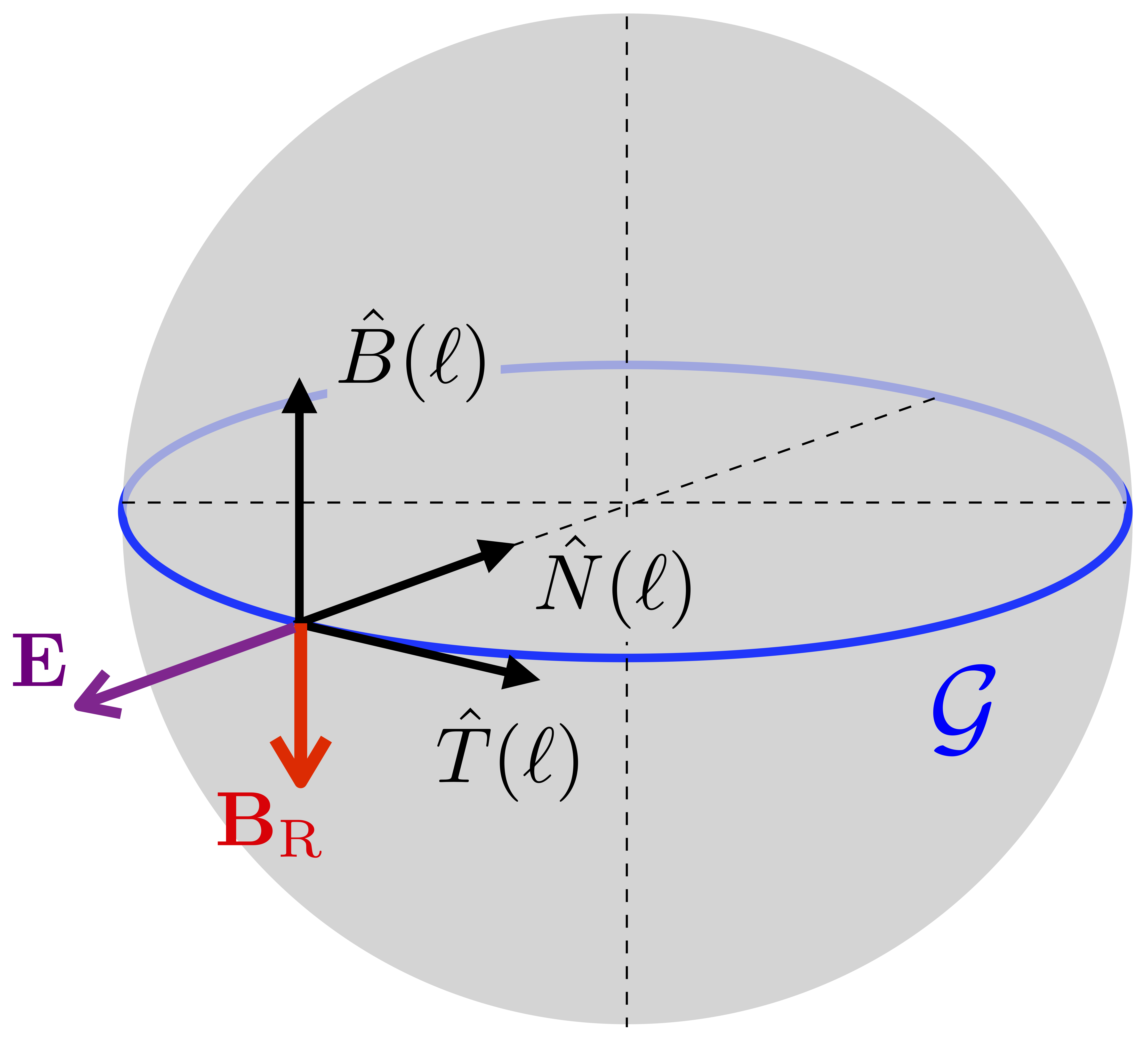}
\caption{\label{geodesic} Geodesic curve $\mathcal{G}$ (great circle) on the surface of a sphere displaying the effective Rashba field $\mathbf{B}_{\text{R}}$ (produced by a radial electric field $\mathbf{E}\propto \alpha_\text{R}$) and the local Frenet-Serret triad $\{\hat{T}(\ell),\hat{N}(\ell),\hat{B}(\ell)\}$, parametrized by the arclength $\ell$: $\hat{N}(\ell)$ points to the sphere's center while $\{\hat{T}(\ell),\hat{B}(\ell)\}$ generate a tangent plane. Importantly, notice that $\hat{B}(\ell)$ and $\mathbf{B}_{\text{R}}$ are constant and antiparallel/parallel for positive/negative Rashba SOC strengths $\alpha_{\text{R}}$.} 
\end{figure}

We define the Rashba sphere of radius $R$ as a curved 2DEG forming a closed manifold of constant and positive Gaussian curvature $1/R^2$, subject to an electric field $\mathbf{E}$ of constant magnitude that points along the radial direction $\hat{R}$. This radial field can be generated by an asymmetric potential confining the 2DEG~\cite{Chang2013} or by surface strain~\cite{Salamone2022}, both formally equivalent to the presence of a central electric charge on the sphere. For geodesic curves $\mathcal{G}$ defined along great circles of radius $R$ we find $\hat{R}(\ell)=-\hat{N}(\ell)$ with vanishing $\alpha_T$ and $\alpha_B$ and constant curvature $\kappa=1/R$. Moreover, $\hat{B}(\ell)$ is constant and tangent to the sphere along $\mathcal{G}$, as shown in Fig.~\ref{geodesic}. In this case, we find that the Hamiltonian~\eqref{H1} reduces to
\begin{equation}
    H_{\mathcal{G}}=-\frac{\hbar^2}{2m}\left( \partial_\ell^2+\frac{1}{4R^2} \right)
    +i\alpha_\text{R} \sigma_B\partial_\ell, \label{H3}
\end{equation}
where we have defined $\alpha_{\text{R}}=-\alpha_N$. The last term in Eq.~\eqref{H3} can be written as $i\alpha_{\text{R}} \sigma_B\partial_\ell=(\mu/2)\mathbf{B}_{\text{R}}\cdot\boldsymbol{\sigma}$, with $\mu$ the Bohr magneton. It corresponds to an effective magnetic field $\mathbf{B}_{\text{R}}$ (proportional to the linear momentum $p_\ell=-i\hbar\partial_\ell$ and antiparallel to $\hat{B}$ for positive $\alpha_{\text{R}}$) acting on the spin carriers as they propagate along the geodesic wire. Notice the difference with usual Rashba rings on flat 2DEGs subject to radial effective magnetic textures, corresponding to a $\mathbf{B}_{\text{R}}$ pointing along $\hat{N}(\ell)$ in Fig. \ref{geodesic}, instead, and leading to rich Aharonov-Anandan geometric phases~\cite{Frustaglia2020}. Indeed, the situation described by Eq.~\eqref{H3} recalls the original proposal by Aharonov and Casher~\cite{AharonovCasher1984} for spin carriers winding an electrically charged line.  

\section{Parallel spin transport}
\label{sec:PST}

A vector field $\mathbf{V}$ tangent to a smooth surface is called \emph{parallel} along the curve $\mathcal{C}$ if its covariant derivative $\nabla$ along $\mathcal{C}$ vanishes, i.e., if $\nabla_{\hat{T}(\ell)}\mathbf{V}(\ell)=0$ with $\ell$ the arclength~\cite{doCarmo2016}. For a geodesic curve $\mathcal{G}$ on a spherical surface, this means that $\mathbf{V}(\ell)=a \hat{T}(\ell)+ b \hat{B}(\ell)$ with constant $a$ and $b$, where the basis $\{\hat{T}(\ell),\hat{B}(\ell)\}$ generates the plane tangent to the sphere along $\mathcal{G}$ as depicted in Fig.~\ref{geodesic}. When considering closed circuits on a curved space, a general geometrical consequence of the curvature is the celebrated concept of \emph{holonomy}: when performing parallel transport, the initial and final orientations of tangent vectors generally differ. 
Consider, e.g., a closed circuit on a sphere that starts at the north pole by following a meridian to the equator, then continues along the equator to the antipodes, and finally returns to the north pole along the corresponding meridian. A tangent vector transported parallelly along this circuit would undergo a holonomy of $180\degree$, with the original and final vectors pointing in opposite directions (e.g., from $\hat{T}$ to $-\hat{T}$). This holonomy coincides with the solid angle $\Omega$ subtended by the circuit from the center of the sphere, which in this example is $\Omega=\Omega_0/4=\pi$ (with $\Omega_0=4\pi$ the solid angle of the full sphere). 

We have so far reviewed the parallel transport of the vector fields transforming under SO(3). As for spinors transforming under SU(2), there are some nuances. Notice that the quantization axis $\hat{\bf n}=\langle \chi|\boldsymbol{\sigma}|\chi\rangle$ of a spinor $|\chi\rangle$ transforms under SO(3). This means that the tangent projection of $\hat{\bf n}$ on the sphere, $\hat{\bf n}_{\parallel}$, responds to the same rules discussed above for a vector field. Moreover, the normal (i.e., radial) projection of $\hat{\bf n}$ on the sphere, $\hat{\bf n}_{\bot}$, is conserved under parallel transport. This indicates that $\hat{\bf n}_{\bot}(\ell)$ coincides with the normal indicatrix $\hat{N}(\ell)$ (up to a constant including a normalization factor and a sign) during parallel transport along a geodesic $\mathcal{G}$. As for the spinor $|\chi\rangle$, aside from the SO(3) holonomy shown by $\hat{\bf n}_{\parallel}$, it undergoes a SU(2) holonomy expressed as an additional phase factor $\text{exp}(-i\Omega/2)$. 

We notice that parallel spin transport can be realized on a Rashba sphere by setting the SOC strength $\alpha_{\text{R}}$ in Eq.~\eqref{H3} to a specific value. This means that the Rashba SOC can act as a geometric connection for spin over the sphere. To show this, we start from the semiclassical expression of the evolution operator $\mathcal{U}$ for a spin carrier propagating along a given path of length $L$
\begin{equation}
\label{Usc}
    \mathcal{U}=\exp{\frac{i}{\hbar}S_{\text{orb}}} U.
\end{equation}
Here, $S_{\text{orb}}=\hbar k_{\text{F}} L$ is the orbital action, with $k_{\text{F}}$ the Fermi wavenumber and $U$ is a unitary operator acting on the spin according to the effective SOC field undergone by the carrier during propagation~\cite{Frisk1993, ZaitsevPRL2005, ZaitsevPRB2005}. From Eq.~\eqref{H3}, along a geodesic curve $\mathcal{G}$ of length $L$ we find 
\begin{equation}
\label{Usc_g}
    U=\exp{ik_{\text{R}}L\sigma_B} 
\end{equation}
with $k_{\text{R}}=\alpha_{\text{R}}m/\hbar^2$ and spin precession length $\lambda_{\text{R}}=\pi/k_{\text{R}}$. The semiclassical expression~\eqref{Usc} holds in the limit $\lambda_{\text{F}}\ll L,\lambda_{\text{R}}$, with $\lambda_{\text{F}}$ the Fermi wavelength. Parallel spin transport is realized by setting $k_\text{R}=-\kappa/2=-1/2R$. In this case, Eq.~\eqref{Usc} reduces to
\begin{equation}
\label{Uscp_p}
    U_\parallel=\exp{-\frac{i}{2}\frac{L}{R}\sigma_B}
\end{equation}
which corresponds to a SU(2) rotation around the $\hat{B}$ axis with angle $\theta=L/R$. To illustrate this, consider the geodesic $\mathcal{G}$ as the equator (the great circle of radius $R$ in the $xy$ plane as shown in Fig.~\ref{geodesic}) and a spinor $|\chi_0\rangle$ placed at $\theta=0$, with $\hat{B}=\hat{z}$ and $\hat{T}=\hat{\theta}$. It is then clear that, for an arclength $L$, $U_\parallel$ applies just the right angle to keep $|\chi_0\rangle$ locally invariant (up to a global phase). For a full round trip with $L=2\pi R$ ($\theta=2\pi$), we find $U_\parallel |\chi_0\rangle =\text{exp}(-i\pi)|\chi_0\rangle = - |\chi_0\rangle$. This shows that the spin quantization axis picks a SO(3) holonomy $\Omega_0/2=2\pi$ (corresponding to the solid angle of a hemisphere) while the spinor itself picks a SU(2) holonomy $\text{exp}(-i\Omega_0/4)$. 
 
\section{Triangular circuits on a Rashba sphere}

\label{sec:TriCircs}

\subsection{Elliptic triangles}

\begin{figure}[!h]
\centering
    \includegraphics[width=0.7\columnwidth]{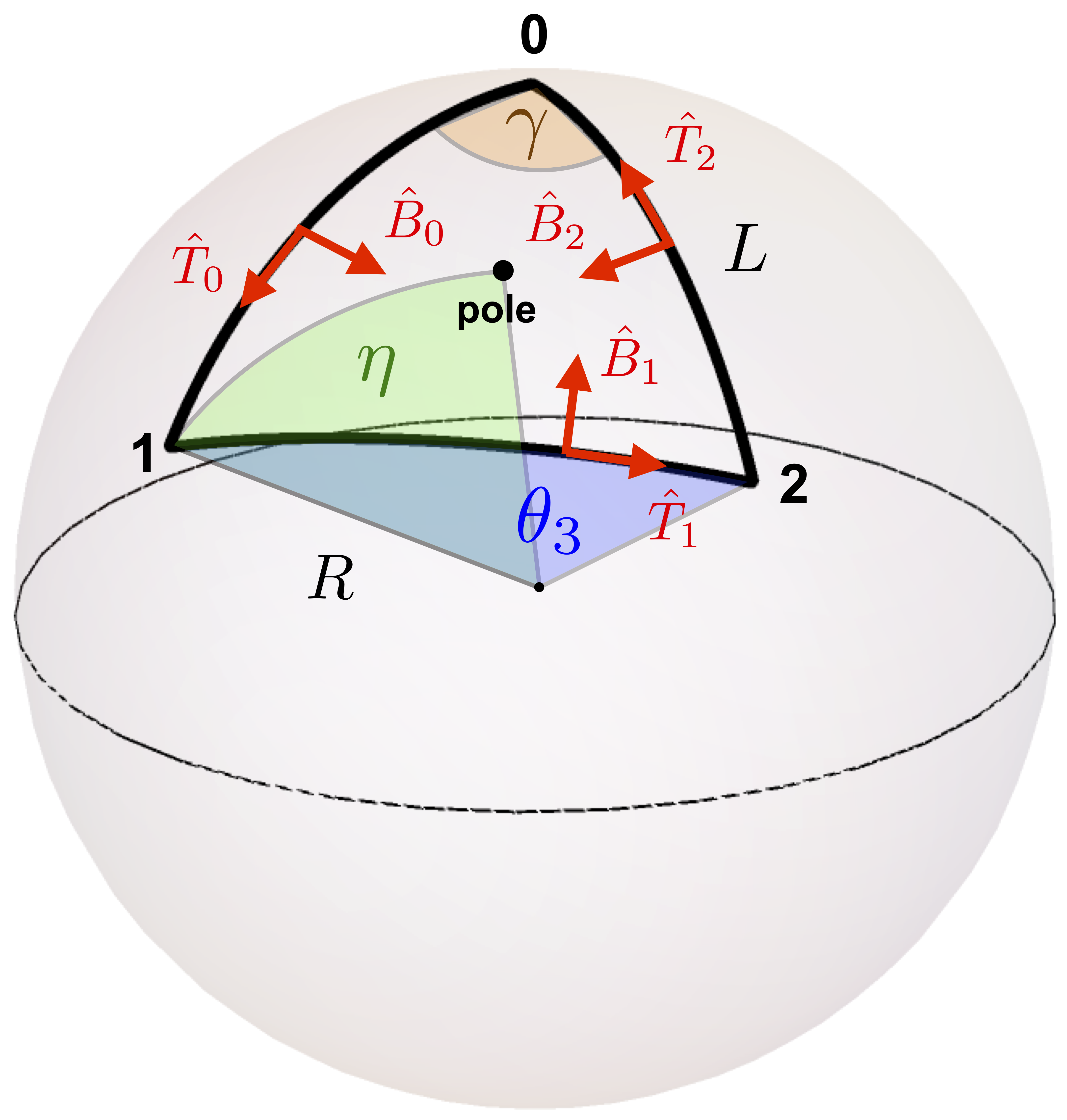}
\caption{\label{triangle} Regular elliptic triangle on a spherical surface of radius $R$ with geodesic sides of length $L$. It is characterized by the polar angle $\eta$, the arc angle $\theta_3=L/R$, and the interior angle $\gamma$. The local tangent bases read $\{\hat{T}_n,\hat{B}_n\}$ with $n=0,1,2$. Notice that $\hat{B}_n$ is constant along the corresponding geodesic segment.} 
\end{figure}

Regular polygons can be defined on the surface of a sphere of radius $R$ by connecting a series of corresponding vertices with geodesic curves | see Fig.~\ref{triangle}. These non-Euclidean polygons are called \emph{elliptic} after the positive Gaussian curvature of the underlying surface. Here, we focus on regular elliptic triangles (elliptic 3-gons), which can be easily generalized to elliptic $N$-gons (see App.~\ref{AppN-gons}). In particular, we consider triangles centered at the north pole with vertices $n=0,1,2$ located at the terminal points of the coordinate vectors $\mathbf{r}_n$ ($|\mathbf{r}_n|=R$). Given the SO(3) rotation matrix 
\begin{equation}
R_{\hat{z}}(2\pi/3) = \begin{pmatrix}
-1/2 & -\sqrt{3}/2 & 0 \\
\sqrt{3}/2 & -1/2 & 0 \\
0 & 0 & 1
\end{pmatrix}\,,
\end{equation}
corresponding to an angle $2\pi/3$ around the $z$ axis, we find $\mathbf{r}_n= [R_{\hat{z}}(2\pi/3)]^n\mathbf{r}_0$ (with $\mathbf{r}_3=\mathbf{r}_0$). For convenience, we choose
\begin{equation}
\mathbf{r}_0=\frac{R}{2}( \sin\eta, -\sqrt{3} \ \sin\eta, 2\cos\eta),
\end{equation}
with $\eta$ the vertices' polar angle. Neighboring vertices $\mathbf{r}_n$ and $\mathbf{r}_{n+1}$ are connected by geodesic arcs $\mathcal{G}_n$ of length $L$. Notice that the unit vector $\hat{B}_n$ along $\mathcal{G}_n$ is such that $\hat{\mathbf{r}}_n \times \hat{\mathbf{r}}_{n+1}=\sin\theta_3 \hat{B}_n$, with $\theta_3=L/R$ and normalized $\hat{\mathbf{r}}_n=\mathbf{r}_n/R$. Besides, $\hat{B}_n=[R_{\hat{z}}(2\pi/3)]^n\hat{B}_0$ due to symmetry. The points $\mathbf{r}^{(n)}(\theta)$ along $\mathcal{G}_n$ are obtained by rotating $\mathbf{r}_n$ an angle $0\le \theta\le \theta_3$ along $\hat{B}_n$, such that 
\begin{equation}
\mathbf{r}^{(n)}(\theta)=R_{\hat{B}_n}(\theta) \mathbf{r}_n
\end{equation}
with $0\le \theta_3 \le 2\pi/3$. Euclidean triangles correspond to the limit $\theta_3\rightarrow 0$ for either small size $L$ or large curvature radius $R$, while for $\theta_3 = 2\pi/3$ the elliptic triangle becomes a great circle, with the vertices lying on the equator ($\eta=\pi/2$). 

It is convenient to write $\hat{B}_0$ in terms of $\theta_3=L/R$ rather than $\eta$. By applying Napier's rules for spherical triangles, we find
\begin{equation}
\label{B_0}
\hat{B}_0=\frac{2}{\sqrt{3}\cos\frac{\theta_3}{2}}\left(-\sqrt{\frac{3}{4}-\sin^2\frac{\theta_3}{2}},0,\frac{1}{2}\sin\frac{\theta_3}{2}\right).
\end{equation}
Moreover, the interior angle between two geodesic sides, $\gamma$, satisfies $2\sin(\gamma/2)\cos(\theta_3/2)=1$. This means that $\gamma$ runs from $\pi/3$ ($\theta_3=0$, corresponding to the Euclidean triangle) to $\pi$ ($\theta_3=2\pi/3$, corresponding to the largest elliptic triangle).

\begin{figure}[!h]
\centering
    \includegraphics[width=0.6\columnwidth]{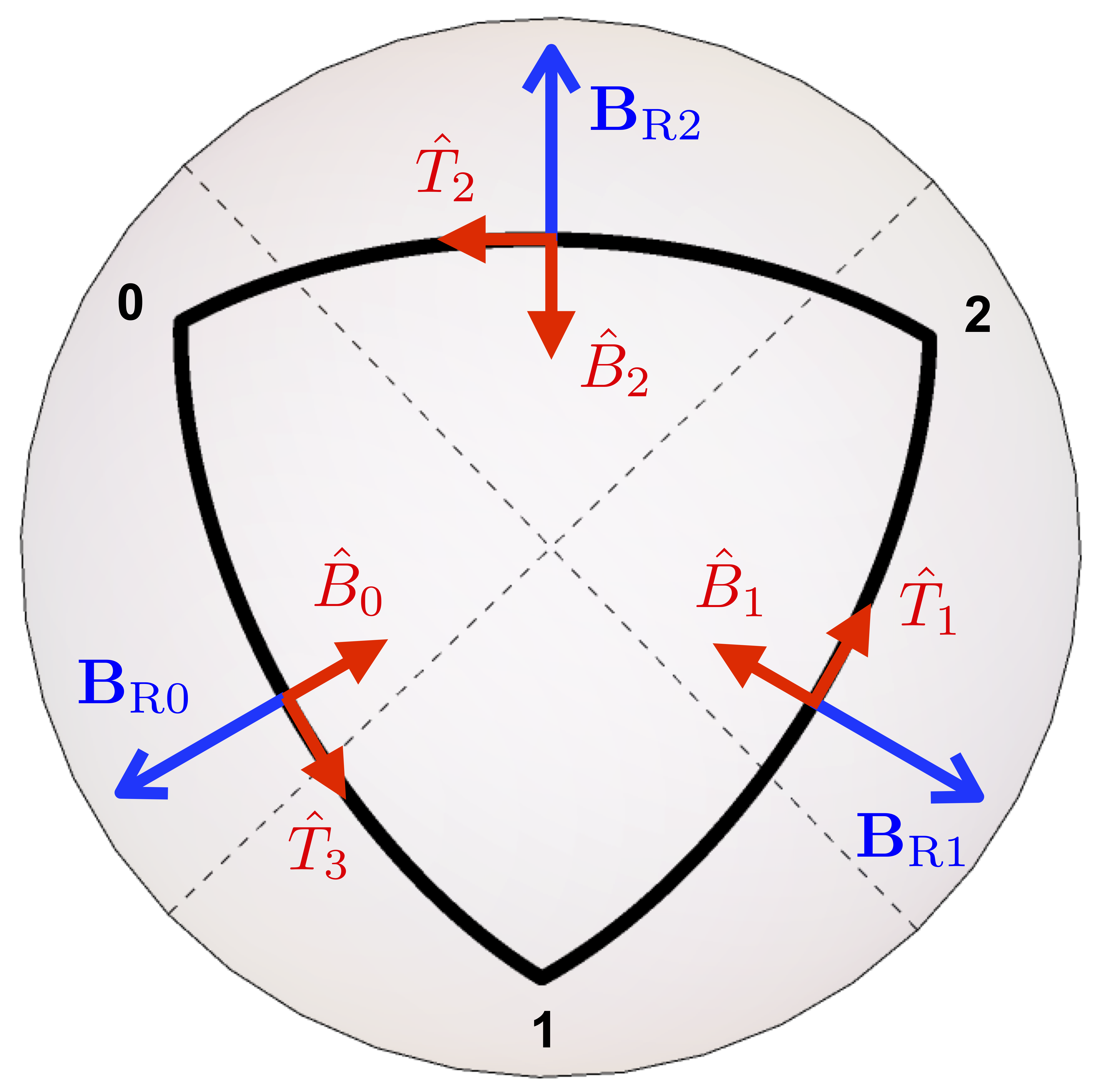}
    \caption{\label{Tcircuit} Top view of a triangular circuit on a Rashba sphere. It displays the effective Rashba field texture $\{\mathbf{B}_{\text{R}n}\}$ (antiparallel to the unit vectors $\{\hat{B}_n\}$). The field texture is tangent to the sphere along the geodesic segments and locally perpendicular to the propagation directions $\{\hat{T}_n\}$. In the Euclidean limit, it reduces to a coplanar texture.}
\end{figure}

It is worth devoting a few words to the role played by the vertices in the carrier dynamics. From the viewpoint of the spin, the discontinuity of $\hat{B}_n$ (and the effective fields $\mathbf{B}_{\text{R}n}$, see Fig.~\ref{Tcircuit}) at the vertices is of utmost importance for the development of complex spin textures and phases~\cite{Liang2025} as a consequence of non-adiabatic spin dynamics discussed in the following sections. These vertices can be safely treated as pointlike discontinuities for Rashba SOC strengths such that the spin precession length $\lambda_{\text{R}}$ is much larger than the vertex size~\cite{Rodriguez2021}. From the viewpoint of the charge, Eq.~\eqref{H1} suggests that the scalar quantum potential for highly curved vertices (with local curvatures $\kappa_n$) creates deep potential wells where the carriers can be localized in their ground state. In the semiclassical limit, such localized potential wells lead to backscattering and Fabry-Perot-like interference effects that contribute to quantum conductance fluctuations. This, however, does not have any consequence neither on the conductance response to spin dynamics nor on the spin phases due to time-reversal symmetry preserved by SOC (as proved in 1D and 2D simulations with polygons~\cite{WangPRL2019, HijanoPRB2021}). Moreover, notice that finite torsions $\tau_n$ arise at the vertices since the elliptic triangular circuit is not a planar curve. However, these local torsions contribute to the scalar quantum potential in the same way the local curvatures $\kappa_n$ do \cite{ortix2015}, and similar arguments apply to their effect on spin-carrier scattering.  

\subsection{Spin phases and curvature}
\label{phases}

From Eq.~\eqref{H3}, we see that the Hamiltonian of propagating spin carriers along each geodesic side of the triangular circuit reads
\begin{equation}
    H_{\mathcal{G}_n}=-\frac{\hbar^2}{2mR^2}\left( \partial_\theta^2+\frac{1}{4} \right)
    +i\frac{\hbar^2k_R}{mR} \sigma_{B_n}\partial_\theta. \label{H4}
\end{equation}
According to Eq.~\eqref{Usc_g}, the spin evolution along ${\mathcal{G}_n}$ is given by 
\begin{equation}
\label{Uscn}
     U_n=\exp{ik_{\text{R}}L\sigma_{B_n}}. 
\end{equation}
Hence, the spin evolution along a full counter-clockwise~(CCW) path around the triangular circuit starting and finishing at vertex $0$ is determined by
\begin{equation}
\label{Uplus}
     U_+=U_2 U_1 U_0. 
\end{equation}
Similarly, spin evolution along clockwise (CW) paths is given by $U_-=U^\dagger_+=U^\dagger_0 U^\dagger_1 U^\dagger_2$ thanks to time-reversal symmetry. This provides a useful tool for studying the spin phases gathered by the carriers around the circuit and modeling the circuit's conductance, following Ref.~\cite{Rodriguez2021}.  

The global AC phase $\phi_s$ gathered by the spin carriers in a round trip is determined through the eigenvalue equation 
\begin{equation}
\label{eigenU}
     U_\pm|\chi_s\rangle=\exp{\pm i\phi_s}|\chi_s\rangle, 
\end{equation}
where $|\chi_s\rangle$ are spinors defined at the initial vertex $0$ with $s=\uparrow, \downarrow$ and $\langle \chi_\uparrow|\chi_\downarrow\rangle=0$. By symmetry, the local spin quantization axis $\hat{\mathbf{n}}_s=\langle \chi_s|\boldsymbol{\sigma}|\chi_s\rangle$ is contained in the plane normal to the sphere that bisects the interior angle $\gamma$. This implies an original misalignment with $\hat{B}_0$ (and $\mathbf{B}_{\text{R}0}$), forcing the spin to precess around $\hat{B}_0$ during propagation along $\mathcal{G}_0$. This repeats identically for every vertex and segment. Consequently, the discontinuity of $\hat{B}_n$ (and $\mathbf{B}_{\text{R}n}$) at the vertices, see Fig.~\ref{Tcircuit}, leads to intricate spin textures in the Bloch sphere as carriers propagate while completing a round trip. In other words, the propagating carriers develop a strongly non-adiabatic spin dynamics \cite{Rodriguez2021}.

The global AC phase $\phi_s$ splits into dynamic, $\phi_{\text{d}}^s$, and geometric, $\phi_{\text{g}}^s$, phase components such that $\phi_s=\phi_{\text{d}}^s+\phi_{\text{g}}^s$. The dynamical spin phase corresponds to the expectation value of the spin Hamiltonian over the propagating spin modes in a CCW round trip. Due to symmetry, this phase reduces to $\phi_{\text{d}}^s=k_{\text{R}}P(\hat{B}_0\cdot\hat{\mathbf{n}}_s)$ with $P=3L$ the triangle's perimeter. The geometric spin phase $\phi_{\text{g}}^s=-\Omega_s/2$, corresponding to a non-adiabatic Aharonov-Anandan phase~\cite{AharonovAnandan87}, is proportional to the solid angle $\Omega_s$ subtended by the spin texture of the propagating modes. The global and dynamic spin phases $\phi_s$ and $\phi_{\text{d}}^s$ are readily obtained by solving Eq.~\eqref{eigenU}, after which the geometric component $\phi_{\text{g}}^s$ is calculated by direct subtraction. 

For simplicity, we focus our discussion on the $s=\uparrow$ spin species defined as the branch for which $|\chi_\uparrow\rangle \rightarrow |z\rangle$ as the Rashba SOC vanishes ($k_{\text{R}}L\rightarrow 0^\pm$). Therefore, from now on, we drop the spin label from the global ($\phi$), dynamic ($\phi_{\text{d}}$), and geometric ($\phi_{\text{g}}$) spin phases. Analytic expressions for these phases and their generalization to $N$-gons on a Rashba sphere can be found in App.~\ref{AppPhases}. 

\begin{figure}[!h]
\centering
    \includegraphics[width=0.85\columnwidth]{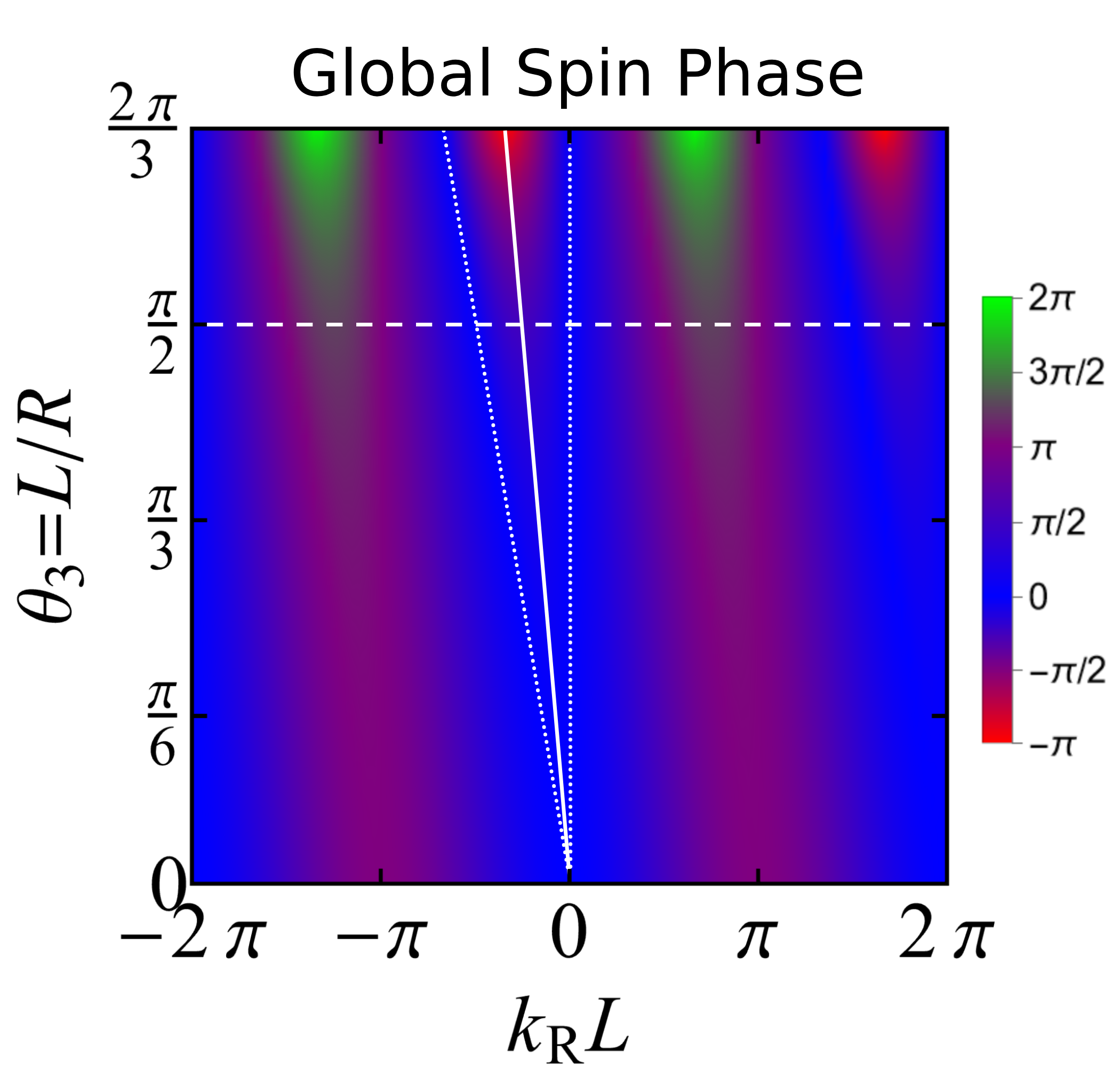}
    \caption{\label{curvaturecont} Response of the global spin phase $\phi$ to the curvature (in terms of $\theta_3=L/R$) and Rashba SOC strength ($k_{\text{R}}L$) in elliptic triangular circuits. The dashed line corresponds to an octant triangle (interior angles $\gamma=\pi/2$). The solid line indicates the parallel spin transport condition. The global spin phase vanishes along the dotted lines ($\phi=0$).} 
\end{figure}

In Fig.~\ref{curvaturecont}, we present our results for the global spin phase $\phi$ as a function of the Rashba SOC strength (in terms of $k_\text{R}L$) and the sphere's curvature (in terms of $\theta_3=L/R$). We find that $\phi$ is bounded and oscillates periodically as a function of $k_\text{R}L$, similarly to what is reported for flat Rashba polygons~\cite{Rodriguez2021}. For the Euclidean triangle ($\theta_3=0$), we find that the periodic pattern is symmetric with respect to $k_\text{R}L=0$. In contrast, Fig.~\ref{curvaturecont} shows that an asymmetric response emerges for finite curvatures: spin carriers propagating in non-Euclidean triangles do not respond symmetrically to the inversion of the Rashba SOC sign as in the Euclidean case. The dashed line in Fig.~\ref{curvaturecont} illustrates the case $\theta_3=\pi/2$ corresponding to a spherical octant (elliptic triangle with interior angles $\gamma=\pi/2$). 
This asymmetric response results as a consequence of parallel spin transport, which introduces an offset indicated by the solid line in Fig.~\ref{curvaturecont} corresponding to the condition $k_{\text{R}}=-1/2R=-\theta_3/2L$. The values taken by $\phi$ along this line coincide with the spin holonomy $-\Omega_3/2$, with $\Omega_3$ the solid angle subtended by the triangular circuit in the spherical 2DEG (see discussion below). The global spin phase $\phi$ turns antisymmetric at $\theta_3=2\pi/3$, with a spin holonomy equal to $-\pi$ (this is not appreciated in Fig.~\ref{curvaturecont} and will be discussed in the following paragraphs). We then conclude from Fig.~\ref{curvaturecont} that the parallel spin transport condition results in a symmetry point for the global spin phase $\phi$, which can be verified plotting $\phi$ as a function of $k'_\text{R}L=(k_\text{R}+1/2R)L=(k_\text{R}+\theta_3/2L)L$.

 Interestingly, the global spin phase $\phi$ vanishes along the dotted lines in Fig. \ref{curvaturecont}. The vertical dotted line on the right corresponds to a vanishing field $k_{\text{R}}L=0$ with trivial phases $\phi=\phi_{\text{d}}=\phi_{\text{g}}=0$ and evolution operators $U_+=U_n=\openone$. The dotted line on the left, instead, is a new branch arising as a consequence of the curvature with finite dynamical and geometric phases $\phi_{\text{d}}=-\phi_{\text{g}}\neq 0$ satisfying $\phi=\phi_{\text{d}}+\phi_{\text{g}}=0$. The same condition appears for general $N$-gons, as further discussed in App.~\ref{AppSC}. Moreover, while $U_+=\openone$ just as in the right branch, the partial operators $U_n \neq \openone$ on the left branch are non-trivial. This results in the development of peculiar spin textures deserving a detailed study beyond the scope of this work. 

\begin{figure}[!h]
\centering
    \includegraphics[width=0.85\columnwidth]{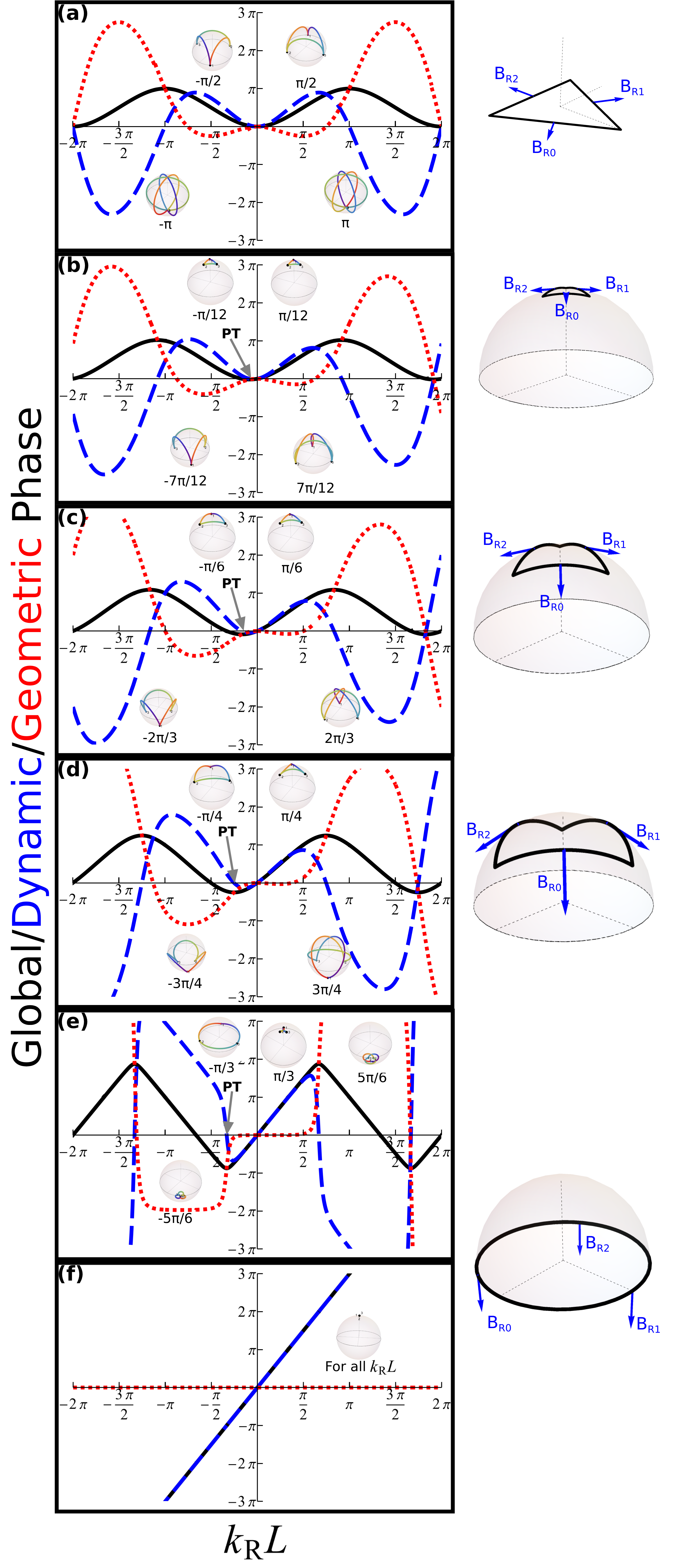}
    \caption{\label{phasesdisc} Global ($\phi$), dynamic ($\phi_{\text{d}}$) and geometric ($\phi_{\text{g}}$) spin phases in elliptic triangles of different curvature $\theta_3=L/R$ as a function of the Rashba SOC strength ($k_{\text{R}}L$): (a) $\theta_3=0$, (b) $\theta_3=\pi/6$, (c) $\theta_3=\pi/3$, (d) $\theta_3=\pi/2$, (e) $\theta_3=2\pi/3-\varepsilon$ with small $\varepsilon$, (f) $\theta_3=2\pi/3$. 
    On the right, sketch of the corresponding triangles and effective Rashba fields. The arrows indicate the parallel spin transport (PT) condition. The insets illustrate the spin textures described in the Bloch sphere by the spin carriers after a round trip for different curvature and Rashba SOC settings. As an example of the asymmetric response to Rashba SOC due to non-Euclidean curvature, notice in (d) the different spin textures and phases displayed at $k_{\text{R}}L=\pm \pi/4$.} 
\end{figure}

\begin{figure}[!h]
\centering
    \includegraphics[width=0.9\columnwidth]{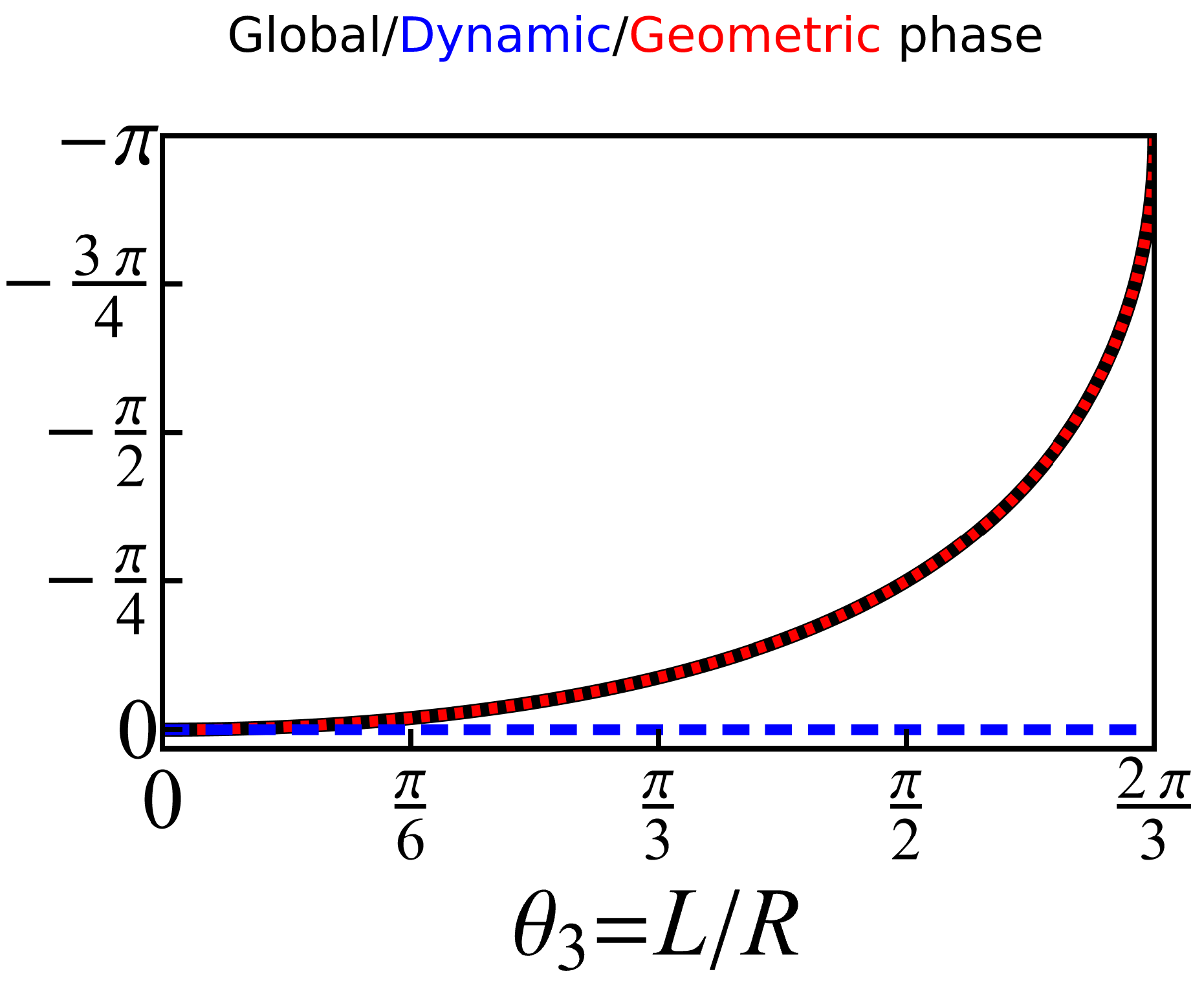}
    \caption{\label{phasesPT} Global ($\phi$), dynamic ($\phi_{\text{d}}$) and geometric ($\phi_{\text{g}}$) spin phases in elliptic triangles along the parallel transport condition ($k_{\text{R}}=-1/2R$, see solid line in Fig.~\ref{curvaturecont}) as a function of the curvature in terms of $\theta_3=L/R$. Notice that here $\phi=\phi_{\text{g}}=-\Omega_3/2$ corresponds to the spin holonomy on the sphere, with $\Omega_3$ the solid angle subtended by the triangular circuit on the spherical 2DEG and by the radial spin texture on the Bloch sphere.} 
\end{figure}

In Fig.~\ref{phasesdisc}, we plot the discriminated spin phases as a function of the Rashba SOC strength ($k_{\text{R}}L$) for different curvatures ($\theta_3=L/R$), from the Euclidean triangle $\theta_3=0$ to the limiting case $\theta_3=2\pi/3$. There, we observe in detail the effect of the curvature on the spin phases. All phases show a nonmonotonic response to the SOC strength, something characteristic of non-circular Rashba circuits \cite{Rodriguez2021}. Interestingly, the global spin phase $\phi$ is bounded by the sum of the interior angles of the circuit such that 
\begin{equation}
\label{phibound1}
\frac{\pi-3\gamma}{2} \le \phi \le \frac{\pi+3\gamma}{2}, 
\end{equation}
with bandwidth $3\gamma$ \footnote{This generalizes to Rashba N-gons such that $[\pi(N-2)-N\gamma]/2 \le \phi \le [\pi(N-2)+N\gamma]/2 $, with bandwidth $N\gamma$, as shown in \cite{Rodriguez2021} for the Euclidean case and in Appendices~\ref{AppPhases} and~\ref{AppSC} for the elliptic non-Euclidean case.}. Moreover, notice that, right at these extremes, the dynamical phase $\phi_{\text{d}}$ vanishes and the global phase is purely geometric (i.e., $\phi=\phi_{\text{g}}$). The first minimum is met at the parallel transport condition (arrows in Fig. \ref{phasesdisc}) | see Fig.~\ref{phasesPT} for a detailed evolution of the spin phases along the parallel transport line. Hence, the global spin phase gathered during parallel spin transport is purely geometric. Moreover, the spin textures are radial, i.e., the spinors $|\chi(\ell)\rangle$ point along $\hat{R}(\ell)$ at every point $\ell$ of the circuit during parallel transport, subtending a solid angle $\Omega_3=3\gamma-\pi$ as a consequence of the Gauss-Bonnet theorem \cite{doCarmo2016}. This suggests to rewrite Eq. (\ref{phibound1}) as
\begin{equation}
\label{phibound2}
-\frac{\Omega_3}{2} \le \phi \le \frac{\Omega_3}{2}+\pi, 
\end{equation}
showing that the lower bound to the global spin phase $\phi$ is nothing but a geometric phase $\phi_{\text{g}}=-\Omega_3/2=(\pi-3\gamma)/2$ corresponding to the SU(2) spin holonomy due to parallel spin transport along the non-Euclidean circuit. This phase appears as a measure of the deviation with respect to the Euclidean case \cite{Dandoloff1992}.

Additionally, the insets in Fig.~\ref{phasesdisc} illustrate how the spin textures respond to positive and negative Rashba SOC. In particular, for $\theta_3=\pi/2$ we notice striking differences at $k_{\text{R}}=\pm1/2R$, with a purely geometric $\phi$ at $k_{\text{R}}=-1/2R$ (open spin texture with large solid angle) and an almost purely dynamic $\phi$ at $k_{\text{R}}=1/2R$ (folded spin texture with small solid angle). Such asymmetries are absent in the Euclidean circuit ($\theta_3=0$). 

The asymmetric response can be addressed from a geometric viewpoint by noticing that CCW propagating spin carriers subject to negative Rashba SOC experience an effective magnetic texture $\mathbf{B}_{\text{R}}$ pointing along $\{\hat{B}_0,\hat{B}_1,\hat{B}_2\}$, in sequential order. For elliptic triangles, this triad is a non-coplanar, right-handed set of vectors, i.e., $(\hat{B}_n\times\hat{B}_{n+1})\cdot\hat{B}_{n+2}>0$. In contrast, positive Rashba SOC produces the left-handed sequence $\{-\hat{B}_0,-\hat{B}_1,-\hat{B}_2\}$, leading to very different spin dynamics and phases. It is only for Euclidean triangles that the vector sequence is coplanar, and the inversion of the Rashba SOC field leads to a symmetric spin response.

The limiting case $\theta_3=2\pi/3$ deserves a few comments. It corresponds to a complete great circle as the one depicted in Fig.~\ref{geodesic}, where the effective Rashba field $\mathbf{B}_{\text{R}}$ is uniform pointing along the $z$ axis. Our choice for the spinor $|\chi_\uparrow\rangle$ is aligned with the $z$ axis in this case. This means that the geometric phase should vanish (since no solid angle is subtended) and the global spin phase should be purely dynamical (i.e., $\phi=\phi_{\text{d}}$) and antisymmetric with respect to $k_{\text{R}}L=0$. This is shown in Fig.~\ref{phasesdisc}(f), as expected. Still, this (anti)symmetry is not observed in Fig.~\ref{curvaturecont} due to the critical response of the phases near $\theta_{3}=2\pi/3$, where virtually no numerical resolution is sufficient to capture that feature. 
This is illustrated in Fig.~\ref{phasesdisc}(e), corresponding to $\theta_3=2\pi/3-\varepsilon$ with small $\varepsilon$. In this situation, the spin carriers experience a dominant effective field along the $z$ direction subject to a small perturbation mixing the two spin species, leading to complex spin textures and phase dependencies as a function of the SOC strength. In the limit $\varepsilon \rightarrow 0$, both descriptions are eventually equivalent.  
\begin{figure}[!ht]
\centering
    \includegraphics[width=0.7\columnwidth]{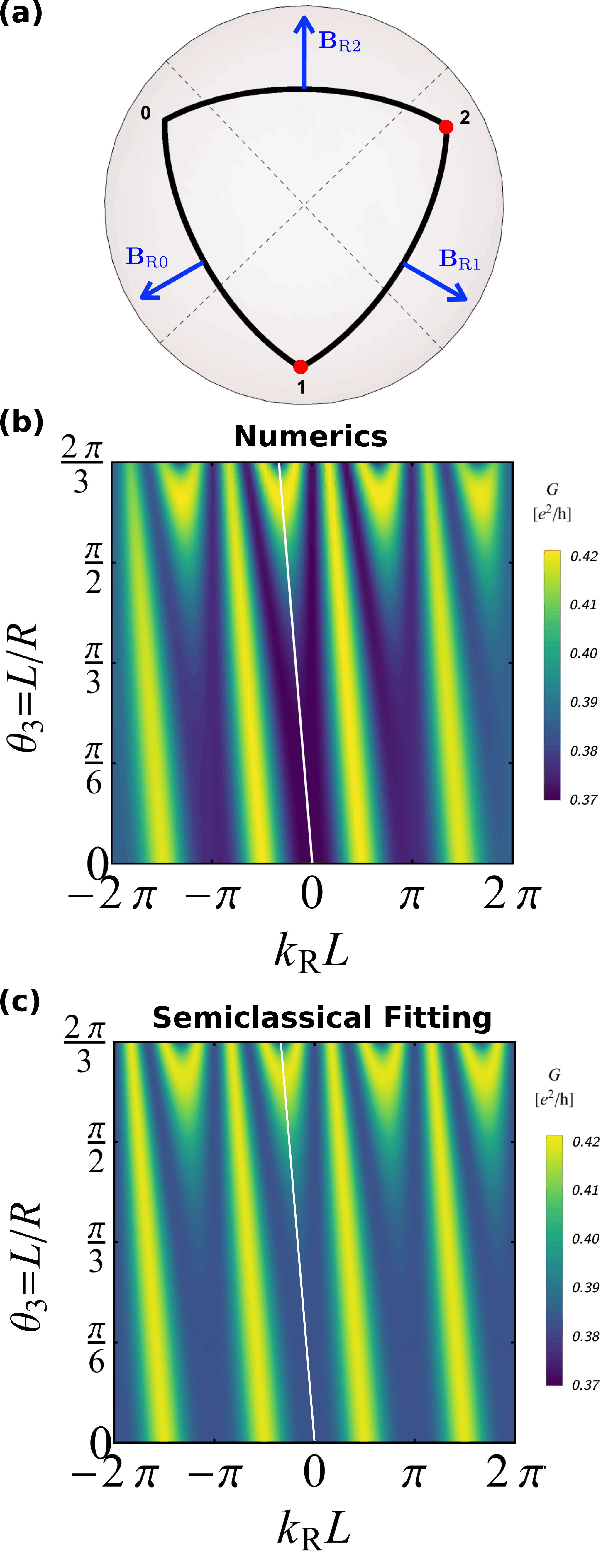}
    \caption{\label{G2} Conductance $G$ of a disordered elliptic triangle in units of $e^2/h$ as a function of the curvature (in terms of $\theta_3=L/R$) and the Rashba SOC strength ($k_{\text{R}}L$): (a) sketch of the transport setup with red dots indicating the source and drain contacts; (b) numerical result based on a tight-binding model; (c) semiclassical fitting $G=(e^2/h)(g_0 + g_1 cos{ 2\phi})$ with $g_0=0.4014$ and $g_1=-0.0183$, and global phase $\phi$ taken from Fig.~\ref{curvaturecont}. The solid lines correspond to the parallel spin transport condition.}
\end{figure}

\subsection{Quantum conductance and curvature}

In this section, we show that the curvature of elliptic Rashba triangles has observable consequences in quantum transport. 
We consider electronic transport between two 1D contact leads placed at, e.g., vertices 1 and 2 in Fig.~\ref{Tcircuit}. 
It has been demonstrated~\cite{MeirGefenEntin-Wohlman89, MathurStone92} that the quantum conductance $G$ of electronic circuits depends symmetrically on the global AC phase $\phi$. In Sec.~\ref{phases}, we have shown that the curvature induces an asymmetric response of $\phi$ to the inversion of the Rashba SOC ($k_{\text{R}}L$) sign. As a consequence, we find that the conductance of elliptic triangles displays similar asymmetries in response to Rashba SOC. 

A realistic modeling of the quantum conductance requires considering the effect of disorder. This also creates the conditions for isolating features that do not depend on the details of a particular circuit \cite{NagasawaPRL2012, Nagasawa2013, WangPRL2019}. 
To this aim, we compute the conductance of elliptic triangles for spin unpolarized incoming carriers by performing full quantum simulations based on a 1D tight-binding model (see App. \ref{AppTB} for details) implemented in the open-source \textsc{kwant} code~\cite{Kwant} that applies a scattering wavefunction formalism. Disorder is introduced through fluctuations in the arc lengths such that $L\rightarrow L'=L(1+\delta l)$, where $\delta l \in [-0.1,0.1]$ is taken randomly from a uniform distribution~\cite{VidalPRB2000, BerciouxPRL2004, BerciouxPRB2005, Bercioux2005, HijanoPRB2021, RodriguezPRB2024}. We average the conductance over 100 realizations of disorder. Additionally, to boost numerical convergence, we performed an energy average of over 100 different values in a window larger than the mean-level spacing of the closed triangle. The semiclassical regime is guaranteed by choosing $L/\lambda_{\text{F}} \approx 60$, with $\lambda_{\text{F}}$ the Fermi wavelength. The results for the average conductance as a function of the Rashba SOC strength ($k_{\text{R}}L$) and the curvature ($\theta_3=L/R$) are presented in Fig.~\ref{G2}(b). 
There, we find an AC interference pattern of quantum origin displaying an asymmetric response to Rashba SOC with respect to the zero-field point ($k_{\text{R}}L=0$) induced by the curvature. It is only for the Euclidean triangle ($\theta_3=0$) and in the limiting case $\theta_3=2\pi/3$ that the conductance verifies a symmetric response under the inversion of the Rashba SOC sign. The solid line in Figs.~\ref{G2}(b) and (c) corresponds to the parallel transport condition, highlighting the shift introduced in the AC pattern. As a result, note that the parallel spin transport condition emerges as a symmetry point for the AC conductance. This can be verified by plotting it as a function of $k'_\text{R}L=(k_\text{R}+1/2R)L=(k_\text{R}+\theta_3/2L)L$.

To gain physical insight into the quantum corrections to the conductance responsible for the AC pattern displayed in Fig.~\ref{G2}(b), we propose a path integral approach in the semiclassical limit | see App.~\ref{AppSC} for details. Semiclassical
theory shows that the leading quantum contributions to the conductance of disordered circuits come from the constructive interference of time-reversed paths of the same length and classical action. The resulting conductance reads
\begin{equation}
\label{Gscfull_main}
G=\frac{e^2}{h}\left(g_0+\sum_{n=1}g_n\cos (2n\phi)\right),
\end{equation}
where, $g_0$ is the (dimensionless) classical conductance originating in path self-pairing, while the sum over $n$ accounts for the quantum correction due to time-reversed-path pairing for carriers winding the circuit $n$ times. The factor 2 in the argument of the cosine function represents a frequency doubling in AC interference, also present in the so-called Al'tshuler-Aronov-Spivak magnetoconductance oscillations of disordered circuits~\cite{AAS}.

Equation~\eqref{Gscfull_main} is the general expression for the classical conductance of 1D loop circuits when all contributing time-reversed paths are taken into account. However, as detailed in App.~\ref{AppSC}, a simplified form of Eq.~\eqref{Gscfull_main} where higher harmonics ($n > 1$) are neglected in favour of the first harmonic ($n=1$) suffices to explain all the relevant physics in Fig.~\ref{G2}(b).
Within this approximation, Eq.~\eqref{Gscfull_main} reduces to
\begin{equation}
\label{Gsc_simplyfied}
G=\frac{e^2}{h}(g_0+g_1\cos2\phi).
\end{equation}
Here, $g_0$ accounts for zero-length paths contributing to initial backscattering at the source contact and self-pairing of single-winding paths, while $g_1$ arises from paired time-reversed single-winding paths.

Figure~\ref{G2}(c) depicts the semiclassical conductance $G$ of elliptic triangles after Eq.~\eqref{Gsc_simplyfied} and the results of
Fig.~\ref{curvaturecont} for the global spin phase $\phi$, as a function of the Rashba SOC strength ($k_{\text{R}}L$) and the curvature ($\theta_3=L/R$). This expression fits the numerical results
of Fig.~\ref{G2}(b) accurately (up to some local extremes) by choosing $g_0=0.4014$ and $g_1=-0.0183$ after quadratic error minimization. In particular, it explains all features observed in the quantum
conductance pattern of Fig.~\ref{G2}(b) in terms of the global AC phase $\phi$ discussed in Sec.~\ref{phases}, such as the general shape and dependence on SOC and curvature (e.g., AC periodicity) and the parallel transport offset introducing a shift in the AC pattern. It is worth to note that the specific values of $g_0$ and $g_1$ do not affect the response of the quantum conductance to SOC and curvature, which is the focus of our interest (see App.~\ref{AppSC}).

\section{Conclusions}

We have shown that spin carrier dynamics and corresponding spin phases in curved 2DEGs are asymmetric with respect to Rashba SOC sign inversion, in contrast to flat 2DEGs. A symmetric response is recovered when considering the parallel spin transport condition as an offset. 
This is demonstrated by studying the development of spin phases in polygonal circuits defined along geodesic curves on 2DEGs on a sphere. There, we noticed that parallel spin transport can be achieved by selecting a Rashba SOC strength that depends on the sphere's curvature in a simple way.
We find imprints of these effects in the quantum conductance of the circuits through AC interference, where disorder appears as a useful resource to isolate universal features from circuit-dependent ones.  

As for physical platforms and relevant experiments, notice that the realization of full Rashba spheres is not strictly necessary: it would be sufficient to consider curved 2DEGs locally shaped as 
spherical caps with SOC characteristics similar to those of the flat 2DEGs used in previous experiments \cite{Koga_2006,NagasawaPRL2012,WangPRL2019} (e.g., micrometer-size ``bumps" on Rashba 2DEGs likely to be developed with current technologies~\cite{Gentile2022,WangReview2020,WeiReview2023}). 
We have estimated that parallel spin transport on an InGaAs-based spherical cap of radius $1\mu$m would require a Rashba SOC strength of the order of $10^{-13}e$V~m, which is well in the range of what is achievable in standard experiments on  flat 2DEGs, usually reaching $10^{-12}e$V~m~\cite{NagasawaPRL2012,WangPRL2019,Nagasawa2013}. 
Regardless of whether the particular non-Euclidean triangular circuits studied in this work are realizable, we note that the effects of parallel transport and holonomy discussed here would be relevant in any future spin-carrier transport experiment on curved surfaces subject to SOC.

Alternatively, the possibility of mapping curved circuits on spherical 2DEGs into Euclidean circuits on flat 2DEGs incorporating curvature, holonomy, and metric by SOC field engineering is worthy of attention~\footnote{E. J. Rodr\'iguez, D. Bercioux, J. P. Baltan\'as, A. A. Reynoso, and D. Frustaglia, in progress (2024).}. A radically different approach would be to simulate the spin dynamics of the carriers by implementing $U_\pm$, Eq.~\eqref{Uplus}, as a succession of single-qubit gates on a quantum computer~\cite{Devitt2016}. In addition, corresponding spintronic circuits in hyperbolic geometries with negative Gaussian curvature deserve a separate study.
 
Finally, note that the implementation of these ideas in Dirac materials, such as graphene~\cite{Huertas2006,BerciouxPRB2023,Chakraborti2024}, could provide a useful tool for simulating relativistic quantum mechanics in curved spacetime using condensed-matter resources.

\acknowledgments

We acknowledge support from Project No. PID2021-127250NB-I00 (e-QSG) funded by the Spanish MICIU/AEI and by ERDF/EU. D.B. acknowledges support from Project No. PID2020-120614GB-I00 (ENACT) funded by the Spanish MICIU/AEI and by ERDF/EU, the Transnational Common Laboratory $Quantum-ChemPhys$ and the Department of Education of the Basque Government through the project PIBA\_2023\_1\_0007 (STRAINER).  A.A.R. acknowledges support from ANPCyT (Argentina) under Grant No. PICT-2020-SERIEA-03123. D.F. thanks L. Chirolli, M. Campisi, S. Jacobsen and G. Mart\'in-V\'azquez for their helpful comments. 

\appendix

\section{Regular polygons on the sphere}
\label{AppN-gons}

The triangle case considered in Section~\ref{sec:TriCircs} can be extended to regular polygons with $N$-sides on the sphere. In the following sections, we derive and present a unified closed result valid for all such polygons at any desired curvature. 

\subsection*{Geometric aspects of elliptic regular polygons.} 
\begin{figure}[!t]
 \begin{center}
    \includegraphics[trim = 0mm 0mm 0mm 0mm,clip=true, keepaspectratio=true, width=0.9\columnwidth]{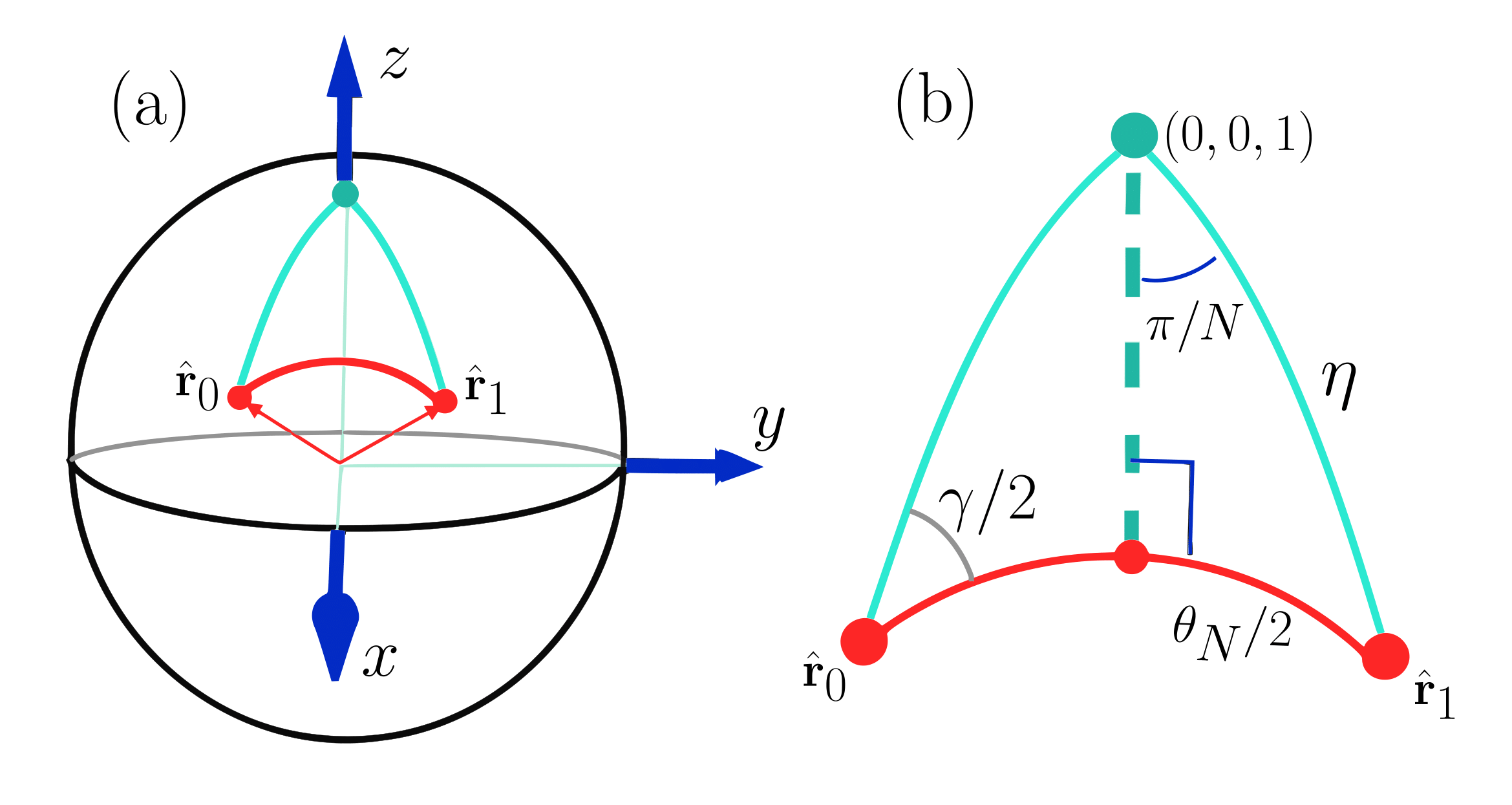}
\end{center}
\vspace{-0.0 cm}

\caption{ Sketch of the first side of a regular-$z$-symmetric $N$-gon on the unit sphere. (a) The geodesic arc $\mathcal{G}_0$ (red line) is the $\theta_N$-angle sector of the great circle that connects $\hat{\mathbf{r}}_0$ and $\hat{\mathbf{r}}_1$. The two vertices lie at a polar angle $\eta$ and, together with the north pole, form a spherical triangle where the angle between the two meridians (green lines) is $2\pi/N$ by construction. (b) Using the bisecting meridian (dashed green line) the spherical triangle of (a) is divided into two equivalent spherical rectangle triangles to which we apply Napier's rules to relate $\theta_N/2$, $\eta$, $\pi/N$, and $\gamma/2$, where $\gamma$ is the interior angle between two consecutive sides.}
\label{FgAp:1}
\end{figure}

In the same fashion as the main text, here the $N$ vertices, labeled with $n\in\{0,1,2,\ldots,N-1\}$, are located at positions $\mathbf{r}_n=R \hat{\mathbf{r}}_n$ with $R$ the sphere's radius. For convenience, as sketched in Fig.~\ref{FgAp:1}(a) for a unit sphere, the polygon has $z$-axis symmetry, with polar angle $\eta$ and $\hat{\mathbf{r}}_0=\left(\cos\frac{\pi}{N}\sin\eta,-\sin\frac{\pi}{N}\sin\eta,\cos\eta\right)
$, i.e., the first side is symmetric with respect to the $x$ axis. The locations of the other vertices are easily generated by $2\pi/N$-angle rotations around the $z$ axis, namely, $\hat{\mathbf{r}}_n= [R_{\hat{z}}(2\pi/N)]^n\hat{\mathbf{r}}_0$,
with $R_{\hat{v}}(\theta)$ defined as the $3\times3$-rotation matrix corresponding to an angle $\theta$  around the direction given by the unit vector $\hat{v}$.

Given two consecutive vertices $n$ and $n+1$ the geodesic arc connecting them, $\mathcal{G}_n$, is defined by a rotation around the binormal direction, $\hat{B}_n=(\hat{\mathbf{r}}_n\times\hat{\mathbf{r}}_{n+1})/\sin\theta_N$, where the angular length of the arc is 
\begin{equation}
    \theta_N\equiv \frac{P}{NR} = \frac{P\kappa}{N}\,,
\label{eq:theN}
\end{equation} 
with $\kappa=1/R$ the curvature and $P$ the perimeter of the polygon. The polygon's positions $\mathbf{r}^{(n)}(\theta)$ along $\mathcal{G}_n$ are obtained by rotating $\mathbf{r}_n$ an angle $0\le \theta\le \theta_N$ around the binormal such that 

\begin{equation}
\mathbf{r}^{(n)}(\theta)=R_{\hat{B}_n}(\theta) \mathbf{r}_n\,.
\label{eq:posiciones}
\end{equation}
The latter expression allows for the obtention of the explicit positions visited as the coordinate $\ell$ (see main text) evolves along the closed path loop.

In order to write the relevant expressions as a function of curvature rather than the polar angle, we resort to Napier's rules applied to the rectangular triangles in Fig.~\ref{FgAp:1}(b), obtaining the relation
\begin{equation}
\sin\frac{\theta_N}{2}=\sin\eta\sin\frac{\pi}{N}\,.
\label{eq:napier_rule}
\end{equation}
From Eqs.~\eqref{eq:theN} and \eqref{eq:napier_rule} it is clear that $\theta_N$ (or $N\theta_N= P\kappa$) quantifies the curvature of the polygon. For example, for $\eta\rightarrow 0$ then $\theta_N\rightarrow 0$ and the polygon is planar, whereas the curvature is maximum for $\eta\rightarrow \pi/2$ as the polygon becomes the equatorial great circle and  $\theta_N\rightarrow 2\pi/N$, i.e., $P\kappa\rightarrow 2\pi$. 

Given our choice of $\hat{\mathbf{r}}_0$ and  $\hat{\mathbf{r}}_1$ the binormal direction for $\mathcal{G}_0$ becomes
\begin{equation}
\hat{B}_0 = \frac{\hat{z}  \cos\frac{\pi}{N}\sin\frac{\theta_N}{2}-\hat{x}\sqrt{\sin^2\frac{\pi}{N} - \sin^2\frac{\theta_N}{2}}}{\sin\frac{\pi}{N}  \cos\frac{\theta_N}{2}}\,.
\label{eq:alphan2}
\end{equation}
Note that due to the $z$-axis symmetry the remaining $\hat{B}_n$ can be obtained from $\hat{B}_0$ by sequential rotations simply as $\hat{B}_n = \left(R_{\hat{z}}\left(\frac{2\pi}{N}\right)\right)^n\cdot \hat{B}_0$. Since the $z$-axis component of  $\hat{B}_n$ does not depend on $n$ it is conceptually useful to define the angle $\vartheta_z$ and write $\hat{B}_0 =\hat{z}\cos\vartheta_z -\hat{x}\sin\vartheta_z$; this angle relates to the others as follows,
\begin{eqnarray}
\tan\vartheta_z   &=& \cot\eta \sec\frac{\pi}{N}\,.\\
\cos\vartheta_z &=& \cot\frac{\pi}{N}\tan\frac{\theta_N}{2} \label{thetaZ}
\,.
\end{eqnarray}
Note that for the planar case, i.e., when $\eta\rightarrow 0$ and $\theta_N \rightarrow 0$, one obtains $\vartheta_z=\pi/2$ and all the $\hat{B}_n$ directions (and thus all their associated $\hat{B}_{\mathrm{R}n}=-\hat{B}_n$ SOC field directions) lie in the $z=0$ plane. As curvature grows, so does the $z$-component $\cos\vartheta_z$, while the $x$- and $y$- components get reduced. It turns out, as discussed in App.~\ref{AppSC}, that $\cos\vartheta_z$ is related to the geometric and dynamic phases at a nontrivial identity robustly appearing at finite values of the spin-orbit field.

Another important feature of the polygon is the interior angle between two consecutive sides, $\gamma$, which appears halved in the rectangular triangles of Fig.~\ref{FgAp:1}(b). In the planar case, the interior angle of an $N$-sided regular polygon is just $\gamma_0=(N-2)\pi/N$. In the spherical case, Napier's rules lead to the relation
\begin{equation}
\cos\frac{\pi}{N}=\sin\frac{\gamma}{2}\cos\frac{\theta_N}{2}~\Rightarrow~\gamma=2\arcsin\left( \frac{\cos\frac{\pi}{N}}{\cos\frac{\theta_N}{2}}\right)  \,.
\label{eq:napier_rule2}
\end{equation}
With curvature, in general,  $N\gamma$, becomes larger than the sum of interior angles of a planar $N$-sided regular polygon: $N\gamma_0 = (N-2)\pi$. The maximum difference appears for the case of $\theta_N=2\pi/N$ where the $N$ vertices lie at the equator $(\eta=\pi/2$ and $P/R=2\pi$) and the polygon becomes a circle, leading to $\gamma=\pi$ and therefore a $2\pi$ excess in the sum of interior angles. By virtue of Girard's theorem, such difference, known as the \emph{excess angle}, is directly related to the area of the spherical $N$-gon, $A_N$, leading to its solid angle expression 
\begin{eqnarray}
\Omega_N\equiv\frac{A_N}{R^2}&=& N\gamma-(N-2)\pi\,, \nonumber \\    &=& 2 N \arcsin\left( \frac{\cos\frac{\pi}{N}}{\cos\frac{\theta_N}{2}}\right)-(N-2)\pi\,.
 \label{eq:solidangle}
\end{eqnarray}
This is a particular case of the Gauss-Bonnet theorem applied to the sphere. The solid angle can be readily put as a function of the perimeter by using $P/R=P\kappa=N\theta_N$. For the case of $N\rightarrow\infty$, the polygon becomes a \emph{ring}, i.e., a constant latitude geographical parallel on the sphere. In such case Eq.~(\ref{eq:solidangle}) tends to 
\begin{equation}
\Omega_{\infty}\equiv 2\pi\left( 1 -\sqrt{1-\left(\frac{P}{2\pi R}\right)^2}\right)\,.
\label{eq:solidangleCAP}
\end{equation}
As expected, the latter expression represents the solid angle of a spherical cap as a function of the cap's base perimeter, being $0$ for $P/R=0$ (planar case) and $2\pi$ for $P/R=2\pi$ (equatorial case). The general $N$ analytic expression for the solid angle, $\Omega_N$, provides the explicit form of the SO(3) and SU(2) holonomies discussed in the main text. In Appendix~\ref{AppPhases}, the connection of such spin holonomy, $-\Omega_N/2$, with the global and geometric phases in parallel transport becomes explicit.

\section{Spin phases in elliptic $N$-gons}
\label{AppPhases}
As mentioned in the main text, the spin evolution along each geodesic arc of length $P/N$ on the Rashba sphere, $\mathcal{G}_n$, reads, according to Eq.~\eqref{Uscn}, 
\begin{equation}
\label{UnAP}
 U_n=\mathrm{exp} \left(\mathrm{i} \sigma_{B_n} k_{\mathrm{R}}P/N\right) \,,
 \end{equation}
 where $\sigma_{B_n} =\hat{B}_n \cdot \boldsymbol{\sigma}$. Such unitary transformation can be interpreted as a $2 k_{\mathrm{R}}P/N$-angle rotation around the $-\hat{B}_n$ direction (which is equal to the main text defined $\hat{B}_{\mathrm{R}n}$ direction), or, equivalently, as a $-2k_{\mathrm{R}}P/N$-angle rotation around the binormal direction $\hat{B}_n$.

\subsection{Parallel transport condition.} The trajectory along the geodesic arc is a $\theta_N$ angle rotation around the $\hat{B}_n$ direction. Therefore, when $\frac{2k_{\mathrm{R}} P}{N}= -\theta_N$ (which is a negative Rashba value) the spin rotation matches the sphere curvature. This condition is known as parallel transport because it ensures that: (i) if the spin starts normal to the sphere at vertex $n$ it evolves being normal to the sphere all along the geodesic arc, to the following vertex, and sequentially, through the full polygon, and (ii) if the spin starts in the plane tangent to the sphere it evolves without leaving such plane. Using that $\theta_N=P/(N R)$ we rewrite this condition as    
 \begin{equation}
     \frac{k_\mathrm{R} P}{N} +  \frac{\theta_N}{2} =0~~~  \Rightarrow~~~  k_\mathrm{R} P +  \frac{1}{2}  ~   \frac{P}{ R}    = 0\,. 
\label{eq:ParaTrans}
\end{equation}
Note that in Eq.~\eqref{eq:ParaTrans}, the perimeter could be factored out, yielding a relation between the spin-orbit coupling strength and the sphere's size: $k_\mathrm{R} R = -1/2$. This demonstrates that the parallel transport condition is independent of the path type over the sphere. However, for our analysis of regular polygons, it is beneficial to retain both variables $k_\mathrm{R} P$ and $P/R$ as in Eq.~\eqref{eq:ParaTrans}. The variable $k_\mathrm{R} P$ relates to (half) the spin-orbit induced rotation angle over one period length. Conversely, $P/R = P\kappa$, as previously noted, quantifies the sphere curvature's effect on the polygon.

The spin evolution along a full CCW path around the $N$-gon circuit, starting and finishing at vertex 0, is determined by
\begin{equation}
\label{UplusAP}
     U_+=U_{N-1} \ldots U_1 U_0\,. 
\end{equation}
To obtain its diagonalization, instead of computing the product of the $N$ different evolution operators in Eq.~\eqref{UplusAP} it is useful to apply symmetry arguments. First, we define the $z$-axis $2\pi/N$-angle rotation in spin space, $Z_N \equiv \mathrm{exp}(-\mathrm{i}\frac{\pi}{N}\sigma_z)$. Due to the polygon rotation $z$-symmetry that links the $\hat{B}_n$ directions it holds that 
\beq
U_{n+1}= Z_N   U_n   Z^\dagger_N\,.
\eeq
\begin{widetext}
We show this explicitly starting from $U_0=\mathrm{exp}\left(\mathrm{i}\frac{k_{\mathrm{R}} P}{N} \hat{B}_0 \cdot \boldsymbol{\sigma}\right)$ and obtaining 
\bea
Z_N   U_0   Z^\dagger_N&=& Z_N  \left(  \cos \frac{k_{\mathrm{R}} P}{N} \openone  -\ci \sin \frac{k_{\mathrm{R}} P}{N}\left(  \sin{\vartheta_z}\sigma_x -\cos{\vartheta_z} \sigma_z \right)\right) \hat{Z}^\dagger_N  \nonumber \\
&=& \cos \frac{k_{\mathrm{R}} P}{N} \openone  -\ci \sin \frac{k_{\mathrm{R}} P}{N}\left(  \sin{\vartheta_z} Z_N \sigma_x  Z_N^\dagger -\cos{\vartheta_z} \sigma_z \right)
\nonumber \\
&=& \cos \frac{k_{\mathrm{R}} P}{N} \openone  -\ci \sin \frac{k_{\mathrm{R}} P}{N}\left(  \cos \frac{2 \pi}{N} \sin{\vartheta_z} \sigma_x  +\sin \frac{2 \pi}{N} \sin{\vartheta_z} \sigma_y  -\cos{\vartheta_z} \sigma_z \right)
\nonumber \\
&=&\mathrm{exp}\left(\mathrm{i}\frac{k_{\mathrm{R}} P}{N} \hat{B}_1 \cdot \boldsymbol{\sigma} \right) = U_1\,.
\eea 
Obviously, since the $n=0$ vertex is not special, this relation is valid for any $n$.

Thus we can rewrite $U_+$ as
\bea
U_+ &=& U_{N-1}  \ldots U_2  U_1  U_0 \nonumber \\
&=& \left( \left(Z_N\right)^{N-1}  U_0  \left(Z_N^\dagger\right)^{N-1}  \right)  \ldots  \left( \left(Z_N\right)^2  U_0 \left(Z_N^\dagger\right)^{2} \right)  \left(Z_N  U_0  Z_N^\dagger \right) U_0 \nonumber \\
&=& \left(Z_N\right)^{N}  \left(Z_N^\dagger  U_0 \right)^{N} \nonumber \\ &=& - \left(Z_N^\dagger  U_0 \right)^{N}
\label{eq:Uplus1} 
\eea
where we used that $\left(Z_N\right)^{N} \equiv \mathrm{exp}(-\mathrm{i} \pi \sigma_z)=-\openone$. Since $Z_N^\dagger U_0$ is a $SU(2)$ operator it can be written as  
\begin{equation}
 Z_N^\dagger  U_0 = \mathrm{exp}(\ci \,\nu\, \boldsymbol{\sigma} \cdot \hat{n}_0) = \cos \nu \openone+\ci \sin \nu ~\,\hat{n}_0 \cdot \boldsymbol{\sigma},            \label{eq:ZdUplusEXP}
\end{equation}
with the quantization axis at vertex $0$ parametrized as $\hat{n}_0= \sin\theta_0 \cos\varphi_0 \ \hat{x}+\sin\theta_0 \sin\varphi_0 \ \hat{y}+\cos\theta_0 \ \hat{z}$.  Importantly, $\nu$ and the angles $\theta_0$ and $\varphi_0$ are to be obtained by inspecting the $SU(2)$ decomposition of $Z_N^\dagger  U_0$. This operator has two eigensolutions, the spin-up or spin-down eigenstates along $\hat{n}_0 \cdot \boldsymbol{\sigma}$ at vertex 0. Using the index $s=\uparrow,\downarrow$ (with $s=1,-1$, respectively, in numeric expressions), the solutions are given by
\beq
\label{chi-plus}
\ket{\chi_{\uparrow}}\equiv\ketLR{\Phi_{\uparrow}(\mathbf{r}_0)} = \left( 
\begin{array}{c}
\cos\frac{\theta_0}{2}\\ 
\mathrm{e}^{\ci\varphi_0} \sin\frac{\theta_0}{2}
\end{array}
\right)~~,~~
\ket{\chi_{\downarrow}}\equiv\ketLR{\Phi_{\downarrow}(\mathbf{r}_0)}=\left( 
\begin{array}{c}
\sin\frac{\theta_0}{2}\\ 
-e^{\ci\varphi_0} \cos\frac{\theta_0}{2}
\end{array}
\right)\,,
\eeq
where, as in the main text, we have also defined the $\ket{\chi_s}$ to denote the solutions at vertex $0$.

Equation~\eqref{eq:Uplus1} implies that the states that diagonalize $Z_N^\dagger U_0$ also diagonalize $U_+$, namely,
\beq
Z_N^\dagger U_0  \ketLR{\Phi_{s}(\mathbf{r}_0)} =  \mathrm{e}^{ \mathrm{i} s \nu} \ketLR{\Phi_{s}(\mathbf{r}_0)}~~\Rightarrow~~ U_+ \ketLR{\Phi_{s}(\mathbf{r}_0)}=\mathrm{e}^{ \mathrm{i}  \phi_s} \ketLR{\Phi_{s}(\mathbf{r}_0)} \,,
\label{eq:ZdU0eig0}
\eeq
where the global phase around the loop can be written in terms of $\nu$ as
\beq
\phi_s=s(N \nu  - \pi)\,.
\label{eq:tot-phase0}
\eeq
 Once the solution at vertex $0$ is obtained, the spinor at other path positions of Eq.~\ref{eq:posiciones}, $\ketLR{\Phi_{s}(\mathbf{r}^{(n)}(\theta))}$ (namely, $\ket{\chi_s(\ell)}$, with $\ell$ the coordinate along the loop) may be obtained by gradually propagating $\ketLR{\Phi_{s}(\mathbf{r}_0)}$ to the next vertex and repeating the process. This is achieved by sequentially applying the partial evolution operators along each geodesic arc,  $U_n(\theta) =\mathrm{exp}({\mathrm{i}\frac{k_R P}{N}\frac{\theta}{\theta_N} \hat{B}_n \cdot \mathbf{\sigma}})$,
where the angle $0\leq\theta\leq\theta_N=2\pi/N$ parametrizes the evolution from vertex $n$ up to intermediate positions towards vertex $n+1$. Then the spin-texture of the solution can be obtained by simply computing mean values of the spin vector around the loop, i.e., $\hat{\boldsymbol{n}}_s(\ell)\equiv \bra{\chi_s(\ell)}\boldsymbol{\sigma}\ket{\chi_s(\ell)}$.

The solutions at $\mathbf{r}_0$ become simpler if we rotate the polygon aligning vertex $0$ with the $x$ axis so that $\mathbf{r}_0=\sin\eta\hat{x} +\cos\eta\hat{z}$ and the spin projection of the solution along the $y$ direction cancels by symmetry arguments ($\varphi_0=0$ and $\hat{n}_0\cdot \hat{y}=0$). Given that in our initial configuration, vertices $0$ and $1$ are rotated $-\pi/N$ and $\pi/N$ with respect to the $x$ direction, respectively, we readily achieve our goal by a $z$-axis rotation of the full polygon an angle $2\pi/(2N)=\pi/N$. The effect of this rotation is equivalent to a simple redefinition of $\hat{B}_0$ as 
\beq
\hat{B}_0 = \frac{\hat{z}\cos\frac{\pi}{N}\sin\frac{\theta_N}{2}-\left(\cos\frac{\pi}{N} \hat{x}+\sin\frac{\pi}{N}\hat{y}\right)\sqrt{\left(\sin\frac{\pi}{N}\right)^2 - \left(\sin\frac{\theta_N}{2}\right)^2} }{\sin\frac{\pi}{N}  \cos\frac{\theta_N}{2}}\,.  \label{eq:a0rot}
\eeq
which leads to the following SU(2) decomposition of $Z_N^\dagger U_0$, 
\bea
Z_N^\dagger U_0   &=&\cos \frac{\pi}{N}\left( \cos \frac{k_{\mathrm{R}} P}{N}  - \sin \frac{k_{\mathrm{R}} P}{N}  \tan \frac{\theta_N}{2}\right)\openone  
 -\ci  \frac{\sin \frac{k_{\mathrm{R}} P}{N} \sqrt{\left(\sin\frac{\pi}{N}\right)^2 - \left(\sin\frac{\theta_N}{2}\right)^2}}{\sin\frac{\pi}{N}  \cos\frac{\theta_N}{2}}  \sigma_x      \nonumber\\
 && +\ci \left(\cos\frac{k_{\mathrm{R}} P}{N} \sin\frac{\pi}{N} + 
 \cos\frac{\pi}{N} \cot\frac{\pi}{N} \sin\frac{k_{\mathrm{R}} P}{N}  \tan\frac{\theta_N}{2}\right) \sigma_z\, .
 \label{eq:opsu2}
 \eea 
This leads to the following cosine and sine of the $\nu$ phase 

\begin{subequations}
\bea
\cos \nu  &=& \frac{\cos \frac{\pi}{N}}{\cos \frac{\theta_N}{2}}\cos\left(\frac{k_{\mathrm{R}} P}{N} +\frac{\theta_N}{2}\right)\,,
\label{eq:cos-nu} 
\\ 
\sin \nu  &=& \sqrt{1- \frac{\cos^2 \frac{\pi}{N}}{\cos^2 \frac{\theta_N}{2}}\cos^2\left(\frac{k_{\mathrm{R}} P}{N} +\frac{\theta_N}{2}\right)}\,.
\label{eq:sin-nu} 
\eea
\label{eq:cos-sin-nu} 
\end{subequations} 
The corresponding eigenvectors can be obtained by noting that Eq.~\eqref{eq:opsu2} reveals the $\hat{n}_0 = \sin\theta_0 \hat{x} + \cos\theta_0 \hat{z}$ of Eq.~\eqref{eq:ZdUplusEXP} as
\bse
\bea
\cos\theta_0 &=&\frac{\cos\frac{k_\mathrm{R} P}{N} \sin\frac{\pi}{N} + 
 \cos\frac{\pi}{N} \cot\frac{\pi}{N} \sin\frac{k_\mathrm{R} P}{N}  \tan\frac{\theta_N}{2}}{\sqrt{1- \frac{\cos^2 \frac{\pi}{N}}{\cos^2 \frac{\theta_N}{2}}\cos^2\left(\frac{k_\mathrm{R} P}{N} +\frac{\theta_N}{2}\right)}},\\
 \sin\theta_0 &=&-\frac{\csc\frac{\pi}{N}  \sec\frac{\theta_N}{2}\sin \frac{k_\mathrm{R} P}{N} \sqrt{\left(\sin\frac{\pi}{N}\right)^2 - \left(\sin\frac{\theta_N}{2}\right)^2}}{\sqrt{1- \frac{\cos^2 \frac{\pi}{N}}{\cos^2 \frac{\theta_N}{2}}\cos^2\left(\frac{k_\mathrm{R} P}{N} +\frac{\theta_N}{2}\right)}}\,.
\\~\nonumber
\eea \label{eq:cos-sin-theta0}
\ese
For the equatorial case $\theta_N=2\pi/N$ and this expression reduces to $(\cos\theta_0,\sin\theta_0)=(1,0)$, which correctly describes the spin-orbit axis being always along the $z$ direction. 

Equations~\eqref{eq:tot-phase0}~and~\eqref{eq:cos-sin-nu}~allow us to obtain an analytical general expression of the global phase valid for all $N$-sided polygons at any curvature and at any spin-orbit strength, namely,
\bea
\phi_s &=&  sN \arccos\left( {\cos \frac{\pi}{N}}{\sec \frac{\theta_N}{2}}\cos\left(\frac{k_{\mathrm{R}} P}{N} +\frac{\theta_N}{2}\right)\right) - s\pi \nonumber \\&=&
-\frac{s}{2}  \left[2N \arcsin\left( {\cos \frac{\pi}{N}}{\sec \frac{\theta_N}{2}}\cos\left(\frac{k_{\mathrm{R}} P}{N} +\frac{\theta_N}{2}\right)\right) -(N-2)\pi \right]
\label{eq:TotalPhase}
\eea
where we used that $\arccos{x}+\arcsin{x}=\pi/2$. For completeness, we also write the global phase as a function of $0\leq P/R=P\kappa \leq 2\pi$ (since  $0\leq\theta_N\leq2\pi/N$ and $P=NR\theta_N$) as
\beq
\phi=\phi_\uparrow = -\frac{1}{2}  \left[2N \arcsin\left( {\cos \frac{\pi}{N}}{\sec \frac{P\kappa}{2N}}\cos\left(\frac{k_{\mathrm{R}} P}{N} +\frac{P\kappa}{2N}\right)\right) -(N-2)\pi \right]
\label{eq:TotPhase2}
\eeq  
\end{widetext}
The latter global phase expression corresponds to the $s = \uparrow$ solution branch, or band, for which $\ketLR{\Phi_{\uparrow}(\mathbf{r}_0)} = \ketLR{z}$ when $k_{\mathrm{R}} P \rightarrow 0^\pm$, irrespective of the value of $\theta_N$. This can be seen because under this condition Eq.~\eqref{eq:cos-sin-theta0} yields $(\cos\theta_0, \sin\theta_0) = (1, 0)$. Additionally, we note that this expression for the global phase is continuous and possesses smooth derivatives. This continuity and smoothness is typically lost if the global phase is confined to a $2\pi$-width zone. In what follows, whenever we write $\phi$, dropping the subindex $s$, we are referring to $\phi_\uparrow$. Of course, considerations as the range, or bandwidth, of $\phi$ can be easily extended to the $s=\downarrow$ branch by recalling that $\phi_\downarrow =-\phi_\uparrow$.   

\subsection{Periodicity in $k_{\mathrm{R}} P$ and bandwidth of the global phase.} By inspecting that $k_{\mathrm{R}} P$ enters solely in the cosine $\cos\left(\frac{k_\mathrm{R} P}{N} +\frac{\theta_N}{2}\right)$, we see that the period in $k_{\mathrm{R}} P$ is $2\pi N$. This periodic behaviour of the global phase is easily understood noting that such a modification in $k_{\mathrm{R}} P$ introduce a $4\pi$-angle change in the rotations of each of the $N$ geodesic arcs of the polygon. Being just an extra even number of full revolutions, they only introduce additional spin-identity operators $\openone$'s leading to identical net $U_n$ operators and identical net $U_+$.

Regarding the values of the global phase in Eq.~\eqref{eq:TotalPhase} for the $s=\uparrow$ branch, it holds that $((N-2)\pi -N\gamma)/2 \le\phi\le ((N-2)\pi +N\gamma)/2$, where we have used the definition of the $N$-gon interior angle $\gamma$ introduced in Eq.~\eqref{eq:napier_rule2}. Interestingly, the extension of allowed $\phi$ values, $N\gamma$ (corresponding to the sum of all interior angles of the polygon), is $(N-2)\pi$ for the planar case and, as curvature grows, it becomes increased by the solid angle of the polygon (the excess angle which is the SO(3) holonomy, $\Omega_N$ given in Eq.~\eqref{eq:solidangle}) achieving the maximum value of $N\pi$ at the equatorial case. These bounds can also be rewritten as 
\begin{equation}
\label{phiboundN}
-\frac{\Omega_N}{2} \le \phi \le \frac{\Omega_N}{2}+(N-2)\pi, 
\end{equation}
thus making an explicit connection of these bounds with the SU(2) holonomy at parallel spin transport.

Note that the extreme values given are obtained for the condition $\frac{k_\mathrm{R} P}{N} +\frac{\theta_N}{2} =m\pi$, being minima (maxima) for even (odd) $m$. The below discussed parallel-transport condition corresponds to $m=0$, while nonzero $m$ cases define solutions that differ from the parallel-transport solution in $2\pi m$-angle spin rotations (each of these $m$ revolutions contributing a spin operator $-\openone$) per geodesic arc. For all these extrema, as we show next, it holds whenever $\partial_{k_\mathrm{R}}\phi=0$, the associated dynamical phase is zero and the total phase becomes purely geometric.

\subsection{Dynamical phase.} 
The Hamiltonian restricted to the spin space is constant along each geodesic arc
\bea
 H_{n,n+1}& =&  -\alpha_\mathrm{R} k_\text{F} \,\, \hat{B}_n \cdot \boldsymbol{\sigma} \nonumber \\
 & =&  -k_{\mathrm{R}}  \hbar v_\text{F} \,\, \hat{B}_n \cdot \boldsymbol{\sigma}
 \eea
with $k_{\mathrm{R}}=m^* \alpha_\mathrm{R} /\hbar^2$. It is convenient to define the full polygon Hamiltonian as  
\beq
H=\sum_{n=0}^{N-1}  H_{n,n+1} =  k_{\mathrm{R}}  \hbar v_\text{F}  {h}_\mathrm{poly}
 \eeq
where we have defined the dimensionless spin operator $h_\mathrm{poly}$ grouping the entire sequence of geodesic arcs. 

The dynamical phase is given by
\bea
\phi^s_{\mathrm{d}}&=&-\frac{1}{\hbar}\int_{\mathrm{poly}}\bra{ \chi_{s}(t)}H(t)\ket{ \chi_{s}(t)}dt
 \nonumber \\
&=&-\frac{1}{\hbar v_\text{F}}\int_{\mathrm{poly}} \bra{ \chi_{s}(\ell)}H(\ell)\ket{ \chi_{s}(\ell)}d\ell \nonumber \\
&=&-k_{\mathrm{R}} P   ~  {\left\langle  \langle     h_\mathrm{poly} \rangle \right\rangle}_{s}. 
\label{eq:dyn:averageh}
\eea
Where we have used that $\ell=v_\text{F} t$ and defined
\beq
 {\left\langle  \langle     h_\mathrm{poly} \rangle \right\rangle}_{s} = -\frac{1}{N}\sum_n \bra{\Phi_{s}(\mathbf{r}_n)} \hat{B}_n \cdot \boldsymbol{\sigma} \ket{\Phi_{s}(\mathbf{r}_n)}. 
 \label{eq:averaged:Bns}
 \eeq
In the last expression, the average along the polygon was reduced to an average over the $N$ vertices. This is a consequence of the conservation of the spin projection along the rotation axis $\hat{B}_n$ for each geodesic arc (from vertex $n$ to $n+1$) spin-evolution.  Furthermore, due to the symmetry of the polygon, all vertices contribute equally, and thus the full summation is $N$ times the contribution of vertex $0$, namely,
 \beq
  {\left\langle  \langle    h_\mathrm{poly} \rangle \right\rangle}_{s}= -\bra{\Phi_{s}(\mathbf{r}_0)} \hat{B}_0 \cdot \boldsymbol{\sigma} \ket{\Phi_{s}(\mathbf{r}_0)} = -s  \hat{n}_0\cdot \hat{B}_0 .
 \label{eq:dynfase0} 
\eeq
Equation \eqref{eq:dynfase0} can be evaluated using Eqs.~\eqref{eq:cos-sin-theta0} and~\eqref{eq:a0rot} generating the dynamical phase.  

We can also obtain the dynamical phase through an alternative, more convenient method. First, note that the global phase---i.e., the phase associated with the eigenvalue of the single polygon evolution operator---is mathematically equivalent to the Floquet quasienergy $\varepsilon$ in a time-periodic system (more specifically, to the phase accumulated after a period, $-\varepsilon T/\hbar$; see, for example, \cite{Reynoso:PRB:2013,Reynoso:NJP:2017}). Then, using standard Floquet theory, we obtain the dynamical phase:
\beq
\phi_{\mathrm{d}}^s
= \frac{s k_{\mathrm{R}} P   \cos{ \frac{\pi}{N} }  \sin{\left( \frac{k_{\mathrm{R}} P}{N} + \frac{\theta_N}{2} \right)} }{ 
\sqrt{ \cos^2{\frac{\theta_N}{2} } - \cos^2{ \frac{\pi}{N}} \cos^2{\left( \frac{k_{\mathrm{R}} P}{N} + \frac{\theta_N}{2} \right)} }},
\label{eq:dyn:phase}
\eeq
directly from the global phase of Eq.~\eqref{eq:TotalPhase}. Here, we have used $ k_{\mathrm{R}}\frac{\partial H}{\partial k_{\mathrm{R}}} = H$ and applied the Hellmann-Feynman (HF) theorem:
\beq
\frac{\partial\phi_{s}}{\partial k_{\text{R}}} = -        \frac{1}{\hbar}\int_{\mathrm{poly}} \bra{ \chi_{s}(t)}\frac{\partial H}{\partial k_{\mathrm{R}} }\ket{ \chi_{s}(t)}dt =\frac{\phi_{\mathrm{d}}^s}{k_{\mathrm{R}}} \,.
\label{eq:HFtheorem}
\eeq

\subsection{Geometrical phase.} It can be obtained directly from the global phase of Eq.~\eqref{eq:TotalPhase} and the dynamical phase of Eq.~\eqref{eq:dyn:phase} considering that  $\phi_s=\phi_{\mathrm{g}}^s+\phi_{\mathrm{d}}^s$. The result is  
\begin{widetext}
\beq
\phi_{\mathrm{g}}^s = -\frac{s}{2}  \left[2N \arcsin\left( {\cos \frac{\pi}{N}}{\sec \frac{\theta_N}{2}}\cos\left(\frac{k_{\mathrm{R}} P}{N} +\frac{\theta_N}{2}\right)\right) -(N-2)\pi \right]
 - \frac{s k_{\mathrm{R}} P   \cos{ \frac{\pi}{N} }  \sin{\left( \frac{k_{\mathrm{R}} P}{N} + \frac{\theta_N}{2} \right)} }{ 
\sqrt{ \cos^2{\frac{\theta_N}{2} } - \cos^2{ \frac{\pi}{N}} \cos^2{\left( \frac{k_{\mathrm{R}} P}{N} + \frac{\theta_N}{2} \right)} }}  \label{eq:GeoPhase}
\eeq
\end{widetext}

\subsection{Phases and the parallel transport condition.} Note that for parallel-transport (PT), $\frac{k_{\mathrm{R}} P}{N} +\frac{\theta_N}{2}=0$, $\cos\left(\frac{k_{\mathrm{R}} P}{N} +\frac{\theta_N}{2}\right)=1$ and the \emph{global phase} in Eq.~\eqref{eq:TotalPhase} becomes the spin holonomy: half the solid angle subtended by the spin-texture along the loop, which becomes equivalent to the one subtended by the real-space elliptic trajectory given in Eq.~\eqref{eq:solidangle}.
This confirms that the global phase is purely geometrical for parallel transport, which is also in agreement with the cancellation of the dynamical phase of Eq.~\eqref{eq:dyn:phase} given that $\sin\left(\frac{k_{\mathrm{R}} P}{N} +\frac{\theta_N}{2}\right)=0$. Of course, the latter dynamical phase cancellation follows trivially from the very definition of parallel-transport condition: as the spin evolves being normal to the sphere, the projection along the spin-orbit axis, which lies along the binormal direction, is zero all along any elliptic polygon.   

Another important property of the PT condition arises in the global phase dependence with $k_{\mathrm{R}} P$ for fixed curvature $P/R =N\theta_N$. Since the spin-orbit strength enters solely inside the cosine $\cos\left(\frac{k_{\mathrm{R}} P}{N} +\frac{P}{2RN}\right)$ then the global phase $\phi(k_\text{R}P,P/R)$ is even with respect to the PT condition that nullifies the argument of the cosine. By defining the distance in $k_\text{R}P$ from the PT condition,  $\delta_{k_\text{R}P}\equiv|k_\text{R}P+\frac{P}{2RN}|$, it is easy to see that
\beq 
\phi_s\left(-\frac{P}{2R} +\delta_{k_\text{R}P},\frac{P}{R}\right) = \phi_s\left(-\frac{P}{2R} -\delta_{k_\text{R}P},\frac{P}{R}\right)\,.
\label{symPT0}
\eeq
Furthermore, this symmetry has observable consequences in the conductance that are discussed in App.~\ref{AppSC}.

It is worth noting that for non-zero curvature, $\frac{P}{R}\neq0$, this symmetry is nontrivial since it only holds for the global phase, in contrast to the planar case of $\frac{P}{R}=0$ for which all the phases possess symmetries with respect to $k_\text{R}P=0$. In particular, the fact that the geometric phase of Eq.~\eqref{eq:GeoPhase} is neither odd nor even with respect to PT implies that the solutions related by Eq.~\eqref{symPT0} for nonzero $\delta_{k_\text{R}P}$ have spin textures that enclose different solid angles. This difference is expected because the spin-evolution along each geodesic arc involves a $2k_\text{R}P/N$ angle rotation around the binormal direction, and such an angle is different for the $k_\text{R}P$ values linked by the symmetry, namely,  
 \beq
 (k_\text{R}P)_\pm \equiv -\frac{P}{2R} \pm \delta_{k_\text{R}P}\,.
 \label{eq:symPT:labels}
 \eeq
 
However, the even parity of the global phase around PT in Eq.~\eqref{symPT0} implies that the derivatives with respect to such condition are odd and then
\begin{widetext}
\beq
\left.\frac{\partial\phi_s}{\partial{k_\text{R}P}}
 \right|_{(k_\text{R}P)_+} = - \left. 
\frac{\partial\phi_s}{\partial{k_\text{R}P}} \right|_{(k_\text{R}P)_-} ~~\Rightarrow~~\left. {\left\langle  \langle    h_\mathrm{poly} \rangle \right\rangle}_{s}\right|_{ 
(k_\text{R}P)_+} = - \left.{\left\langle  \langle    h_\mathrm{poly} \rangle \right\rangle}_{s} \right|_{ 
(k_\text{R}P)_-} \,. 
\label{symPTderiv}
\eeq
\end{widetext}
where we have used the HF theorem of Eq.~\eqref{eq:HFtheorem}. Equation~\eqref{symPTderiv} indicates that the spin-textures of the solutions related by the symmetry around the PT have opposite average spin projections of Eqs.~\eqref{eq:averaged:Bns} and ~\eqref{eq:dynfase0}. In App.~\ref{AppSC} we further discuss how this quantity behaves for the important particular case of $\delta_{k_\text{R}P}=\frac{P}{2R}$ in which $(k_\text{R}P)_+=0$ (the trivial identity) and $(k_\text{R}P)_-=-P/R$ (a non-trivial identity related by the symmetry around PT).    

\subsection{The ring or parallel.} Taking the limit $N\rightarrow\infty$ in our expressions yields the solution for a ring (in the planar case). For nonzero curvature, this case is equivalent to a geographical parallel of constant latitude angle $\varphi_l=\pi/2-\eta$. The perimeter of this parallel is $P=2\pi R\cos\varphi_l$, thus the curvature is quantified by $P/R=2\pi\cos\varphi_l$. Note that, unlike the finite $N$ case, here the latitude angle $\varphi_l$ of the vertices is identical to the $\vartheta_z$ angle (as Eq.~\eqref{thetaZ} leads to $\cot\eta\sec\frac{\pi}{N}\rightarrow\tan\varphi_l$). As the electron traverses a parallel, the $z$ components of the $\hat{B}_n$'s remain constant while their $x$-$y$ components continuously rotate around the $z$-axis. Specifically, for vertex $0$ at position $R\hat{\boldsymbol{r}}_0$ with $\hat{\boldsymbol{r}}_0=\sin\eta
\hat{x}+\cos\eta\hat{z}=\cos\varphi_l
\hat{x}+\sin\varphi_l\hat{z}$, the binormal vector is $\hat{B}_0=-\sin\varphi_l
\hat{x}+\cos\varphi_l\hat{z}$. This binormal smoothly rotates around the $z$-axis as the electron advances along the parallel. An alternative approach involves starting from the circular path, transforming to the rotating frame where the problem becomes exactly solvable, then performing the inverse transformation to return the solution to the original frame. This method yields the same result as taking the large $N$ limit of our general $N$ analytic expressions.

We write down the resulting solution for this particular setup. The spinor at vertex $0$ can be cast to Eq.~\eqref{chi-plus} with $\varphi_0=0$ and 

\[\tan{\theta_0}=\frac{k_\text{R}P\sin\varphi_l}{\pi+k_\text{R}P\cos\varphi_l}
=\frac{k_\text{R}P\sqrt{1-\left(\frac{P}{2\pi R}\right)^2}}{\pi+k_\text{R}P \frac{P}{2\pi R}}\,.
\]

The associated global phase becomes
\beq
\phi_{s}=s\left(\sqrt{\pi^2+k_\text{R}P\left(k_\text{R}P+\frac{P}{R}\right)}-\pi\right)
\label{eq:tot:parallel}
\eeq
This expression contains the features discussed in the Fig.~\ref{fig:apC} below, namely, the parallel-transport condition at $k_\text{R}P=-\frac{P}{2R}$ and the nontrivial identity $k_\text{R}P=-\frac{P}{R}$. As expected, the associated dynamic phase,
\beq
\phi_{\text{d}}^{s} = \frac{s k_\text{R}P (2k_\text{R}P + \frac{P}{R})}{2\sqrt{\pi^2 + k_\text{R}P(k_\text{R}P + \frac{P}{R})}}\,,
\label{eq:dyn:parallel}
\eeq
cancels at the PT condition, which is consistent with the spin of the solution evolving normal to the sphere. Indeed, the spin texture of the $s=\uparrow$ solution is identical to the real-space parallel. Hence, the total phase at PT becomes purely geometric, i.e., $\phi_{s}=\phi^{s}_{\mathrm{g}}=-s\Omega_\infty/2$ with $\Omega_\infty$ the solid angle of the spherical cap given in Eq.~\eqref{eq:solidangleCAP}.      

\section{Tight-binding model}\label{AppTB}

As mentioned in the main text, we employ a 1D tight-binding model to perform the full quantum simulations for elliptic triangles illustrated in Fig.~\ref{G2}(b). To achieve this, we make use of the geometric configuration for the $N$-gons as defined in App.~\ref{AppN-gons}. Therefore, each segment of the $N=3$ geodesic triangle, denoted by $i\in\{0,1,2\}$, is influenced by a spin-orbit field oriented antiparallel to the binormal direction  $\hat{B}_i$ (refer to Eq.~\eqref{B_0}, Eq.~\eqref{eq:alphan2}, and Fig.~\ref{Tcircuit}). Each segment is partitioned into $N_s + 1$ sites, represented by $j = \{0, 1, . . . , N_s\}$, with a lattice spacing of $a_0 = \frac{P}{3 N_s}$. By applying the finite difference method to each segment Hamiltonian in Eq.~\eqref{H4}, the following 1D tight-binding Hamiltonian is derived:
\begin{widetext}
\begin{equation}
\label{HTB}
\hat{\mathcal{H}}_{i}=2 t_\mathrm{h}
\sum_{\sigma,j} \hat{c}_{j\sigma}^\dagger\hat{c}_{j\sigma}   -\sum_{j=0}^{N_s-1}\sum_{\sigma\sigma'}\left[ \left( t_\mathrm{h} \sigma_0 +\ci t_\mathrm{R} \hat{B}_{i}(\theta_3)\cdot\boldsymbol{\sigma} \right)_{\sigma\sigma'}\hat{c}_{j+1,\sigma}^\dagger\hat{c}_{j\sigma'}^{}+h.c. \right]\, 
\end{equation}
with $\sigma_0$ the SU(2) identity matrix and $\boldsymbol{\sigma}$ the vector of Pauli matrices. The operators $\hat{c}_{j\sigma}^{}$ and $\hat{c}_{j\sigma}^\dag$ are the annihilation and creation operators, respectively, for an electron located at site $j$ with spin $\sigma=\{ \uparrow,\downarrow \}$ oriented along the $z$ axis. The hopping energy is given by $t_\mathrm{h}=\hbar^2/(2 m^* a_0^2)$, and the Rashba hopping energy is defined as $t_\mathrm{R}=k_\text{R}Lt_\mathrm{h}/N_s$. In Eq.~\eqref{HTB}, we made explicit the dependence of the binormal direction on the curvature $\theta_3\in [0,2\pi/3]$ of the modeled elliptic triangle. 

\end{widetext}

The Hamiltonian for the entire elliptic triangle is the sum of the Hamiltonians for each segment ($\hat{\mathcal{H}}_{p}=\sum_{i=0}^{i=2}\hat{\mathcal{H}}_{i}$), ensuring that diagonal energy terms at vertices are not counted twice.

The system is connected to a source contact lead and a drain contact lead, denoted as $\eta=\mathrm{s},\mathrm{d}$, which are modeled by semi-infinite tight-binding chains with a hopping energy parameter of $t_\eta=2.5t_\mathrm{h}$, to simulate
contact leads with a broad energy bandwidth. The intermediate spin-independent hopping operators, connecting each lead to a specific vertex of the triangle, are set to the average of the hopping energies in the lead and the triangle, $(t_\eta+t_\mathrm{h})/2$, which is a standard choice that minimizes scattering. We calculate the transport properties using the open-source package \textsc{kwant}~\cite{Kwant}. This method first solves for the scattering matrix of the system attached to the mentioned semi-infinite leads. The resulting transmission probabilities are then used to compute the conductance through the Landauer-B\"{u}ttiker formula, $G = (e^2/h)T$.

The simulations depicted in Fig.~\ref{G2}(b) of the main text are conducted as follows. We set the Fermi energy $E_\text{F}=0.2086t_\mathrm{h}$ corresponding to $\lambda_\text{F}/a_0\approx 13.75$, which ensures that the Fermi wavelength is well-resolved by the discretization. Additionally, we set $N_s = 850$, this choice is consistent with the semiclassical regime as the perimeter, $P = 2550a_0$, is significantly larger than the Fermi wavelength, $P/\lambda_\text{F} \approx 185$. 

As mentioned in the main text, we conduct a Fermi energy average of the conductance to remove variations caused by resonance due to finite-size effects. We define the range of the Fermi wavevector as $k_\text{F} \in [2\pi(n_0-2.5)/P, 2\pi(n_0+2.5)/P ]$, where $n_0 = 185$, and carry out the average over 100 realizations within its associated energy range, thus ensuring suppressing fluctuations on the scale of the orbital level spacing. Additionally, we calculate the average conductance over 100 different disorder realizations. Each disorder configuration is generated by the method outlined in the main text, where the length of each side arc is randomly modified using a $\delta l \in [-0.1,0.1]$. This $\delta l$ is drawn from a uniformly distributed probability function, resulting in changes to the arc lengths such that $L\rightarrow L'=L(1+\delta l)$~\cite{VidalPRB2000, BerciouxPRL2004, BerciouxPRB2005, Bercioux2005, HijanoPRB2021, RodriguezPRB2024}.

\section{Semiclassical conductance}
\label{AppSC}

The Landauer-B\"uttiker formulation \cite{Buttiker1985} identifies the two-contact linear conductance $G$ with the quantum transmission and reflection as
\begin{equation}
\label{LB}
G=\frac{e^2}{h} \text{tr}\left[\mathbb{T}\mathbb{T}^\dagger\right]=\frac{e^2}{h} \text{tr}\left[\openone-\mathbb{R}\mathbb{R}^\dagger\right],
\end{equation}
with $\mathbb{T}=[t_{mn}]$ and $\mathbb{R}=[r_{mn}]$, where $t_{mn}$ and $r_{mn}$ are the quantum transmission and reflection amplitudes from incoming ($n$) to outgoing ($m$) modes. The trace of $\openone$ equals the number of available modes. One-dimensional circuits support one single orbital mode and two spin channels. Hence, the indices $n$ and $m$ run on spin, only, and $\openone$ is a $2\times2$ identity matrix. A semiclassical model~\footnote{For reviews on semiclassical theory, see e.g. K. Richter, \href{https://doi.org/10.1007/BFb0109634}{{\it Semiclassical Theory of Mesoscopic Quantum Systems}} (Springer, Berlin, 2000); R.A. Jalabert, in \href{https://doi.org/10.3254/978-1-61499-228-8-145}{{\it New Directions in Quantum Chaos}}, G. Casati, I. Guarneri, and U. Smilansky, eds. (IOS Press, Amsterdam, 2000).} of $G$ for 1D Rashba loops can be developed whenever the carriers wavelength is much smaller than the system size and the spin splitting is much smaller than the kinetic energy (so that the spin dynamics does not alter the orbital one)~\cite{Frisk1993}, in agreement with usual mesoscopic experimental conditions~\cite{NagasawaPRL2012, Nagasawa2013, WangPRL2019}. In this way, by following a path-integral approach and taking the semiclassical limit~\cite{Loss1992}, the quantum transmission and reflection amplitudes can be expressed as
\begin{eqnarray}
\label{tmn}
t_{mn}&=&\sum_\Gamma a_\Gamma e^{ik_{\text F}L_\Gamma} \langle m | U_\Gamma | n\rangle, \\
\label{rmn}
r_{mn}&=& \sum_\Gamma b_\Gamma e^{ik_{\text F}L_\Gamma} \langle m | U_\Gamma | n\rangle,
\end{eqnarray}
namely, as a sum of phase contributions over different classical paths $\Gamma$ of length $L_\Gamma$ that take the spin carriers from entrance to exit, leading to different statistical weights $a_\Gamma$ and $b_\Gamma$, which eventually result in quantum interference. Within this picture, charge and spin contributions are clearly differentiated. The charge contributes with the orbital phase $\text{exp}[ik_{\text F}L_\Gamma]$. As for the spin, carriers entering the system with spin $n$ can leave it with spin $m$ according to the path-dependent spin evolution operator $U_\Gamma$, which is determined by the particular fields experienced by the spin carriers along the classical path. As for the quantum transmission and reflection, they consists of probability terms of the form
\begin{eqnarray}
\label{Tmn}\\
\nonumber
|t_{mn}|^2&=&\sum_{\Gamma,\Gamma'} a_\Gamma a_{\Gamma'}^*e^{ik_{\text F}(L_\Gamma-L_{\Gamma'})} \langle m | U_\Gamma | n\rangle  \langle m | U_{\Gamma'} | n\rangle^*, \\
\nonumber
|r_{mn}|^2&=&\sum_{\Gamma,\Gamma'} b_\Gamma b_{\Gamma'}^*e^{ik_{\text F}(L_\Gamma-L_{\Gamma'})}  \langle m | U_\Gamma | n\rangle  \langle m | U_{\Gamma'} | n\rangle^*.\\
\label{Rmn}
\end{eqnarray}
For a realistic modeling of the experimental conditions, the effects of disorder and/or sample averaging need to be taken into account. This means that the sums in~\eqref{Tmn} and~\eqref{Rmn} need to run over different configurations, including classical path fluctuations. Moreover, an average over a small energy window around the Fermi energy can be also implemented to take into account the effects of finite (but low) temperatures. Due to the presence of the orbital-phase factors $\text{exp}[ik_{\text F}(L_\Gamma-L_{\Gamma'})]$, the averaging procedure shows that the only surviving terms in~\eqref{Tmn} and~\eqref{Rmn} are those corresponding to pairs of paths $\{\Gamma,\Gamma'\}$ with the same geometric length, $L_\Gamma = L_{\Gamma'}$. Other contributions simply average out due to rapid oscillations of the orbital-phase factors. However, identifying these pairs of paths contributing to the transmission in~\eqref{Tmn} is generally difficult unless two-fold reflection symmetry along the axis connecting the contact leads is preserved. Otherwise, it results most convenient to resort to the quantum reflection~\eqref{Rmn} by taking advantage of unitarity. 
Here we notice that for any backscattering path $\Gamma$ contributing to~\eqref{rmn} there exists another path $\tilde{\Gamma}$ with exactly the same length that follows the trajectory defined by $\Gamma$ but in opposite direction. Namely, $\Gamma$ and $\tilde{\Gamma}$ are paired time-reversed paths contributing to~\eqref{Rmn}.

The quantum contribution to the conductance of disordered circuits arises from the interference of paired time-reversed paths winding the circuit, leading to AC oscillations. The simplest model considers incoming spin carriers that enter the circuit without undergoing backscattering at the source contact and 
interfere after only one single full winding. This corresponds to statistical weights $b_1=b_{-1}=1/2$ for CCW/CW single-winding paths of length $P$ and $b_w=0$ for any other path with winding number $w\neq \pm 1$ in Eq.~\eqref{rmn}. Previous works \cite{Frustaglia2004, Nagasawa2013, WangPRL2019} have shown that this model is sufficient to capture the most relevant features of the quantum conductance such as the general shape and SOC dependence of the interference pattern.
\begin{figure*}[!th]
\centering
 \includegraphics[width=0.82\textwidth]{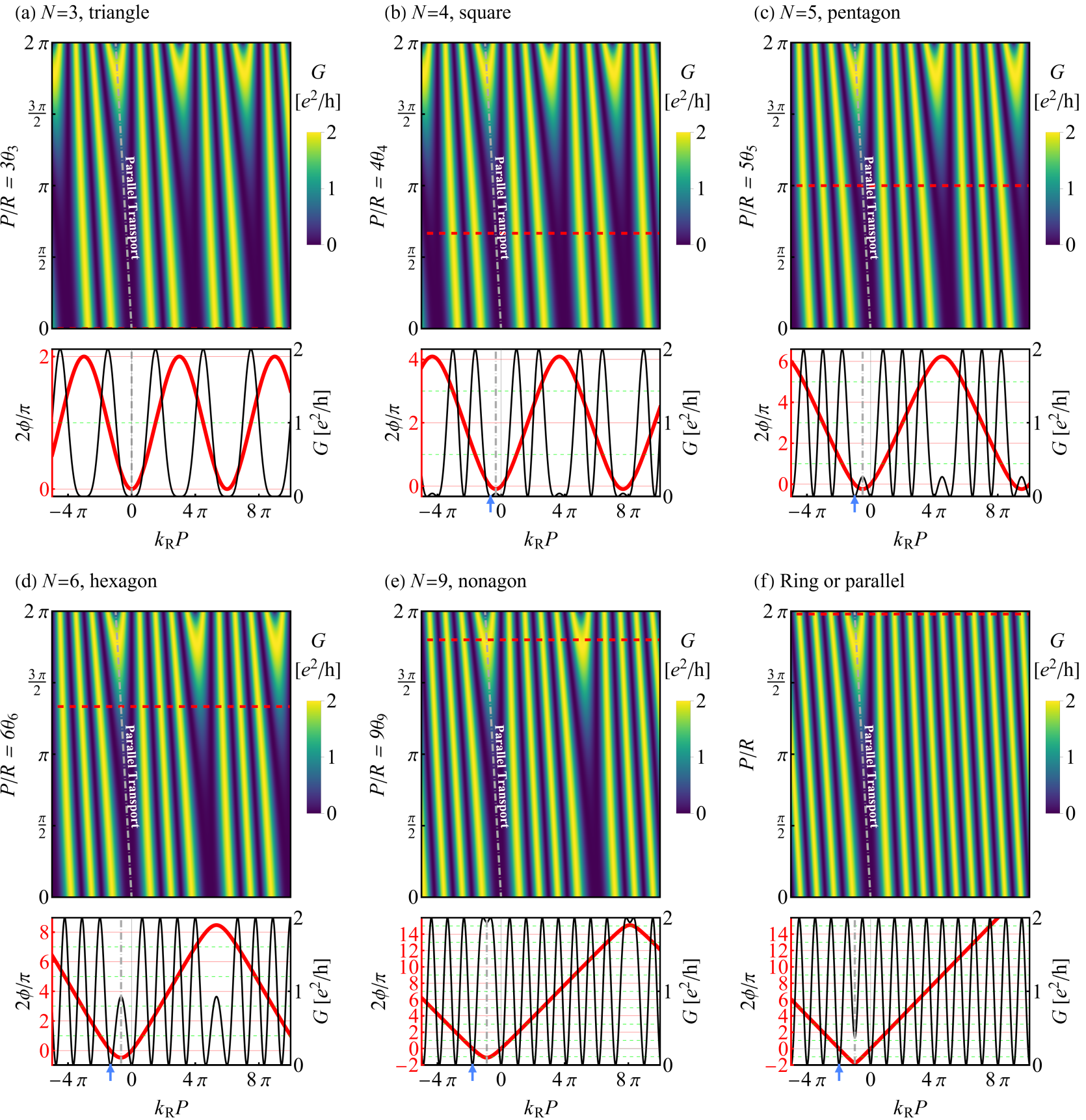}  \caption{Semiclassical conductance $G$ of Eq.~\eqref{Gsc} as a function of $k_\text{R}P$ and the curvature $P/R$ for the case of the (a) triangle, (b) square, (c) pentagon, (d) hexagon, (e) nonagon, and (f) ring (i.e., geographical parallel). The conductance structure near the parallel transport condition, gray-dashed line, is similar for all cases. Each case includes a panel with the $k_\text{R}P$ dependence for $G$ and global phase $\phi$ along the cuts of constant curvatures signalled with horizontal-red-dashed lines in the corresponding $G$ map. Note that as the global phase $\phi$ crosses even (odd) multiples of $\pi/2$ the conductance---proportional to $1-\cos(2\phi)$---realizes a minimum (maximum). For each cut case at increasing curvatures---from (a) to (f) $P/(2\pi R)=\{0,\frac{1}{3},\frac{1}{2},\frac{2}{3},\frac{9}{10},\frac{99}{100}\}$---the blue arrows mark the evolution of the position at which the symmetry around PT of the global phase implies a minimum of the conductance (see text).\label{fig:apC}}
\end{figure*} 
The corresponding reflection amplitudes take the form
\begin{equation}
\label{rmn0}
r_{mn}=\frac{1}{2} \langle m |U_++U_-| n\rangle,
\end{equation}
with $U_+$ defined in Eq.~\eqref{Uplus} for triangular circuits. Notice that in Eq.~\eqref{rmn0} we have dropped a phase prefactor $\text{exp}[ik_{\text F}P]$, irrelevant to the reflection~\eqref{Rmn}. By using time-reversal symmetry ($U_-=U_+^\dagger$) and working in the eigenbasis of $U_\pm$, Eq.~\eqref{eigenU}, we find from Eq.~\eqref{LB} that the conductance within this approximation takes the form 
\begin{equation}
\label{Gsc}
G=\frac{e^2}{h}(1-\cos 2\phi),
\end{equation}
where the minus sign and the factor 2 in the argument are the consequences of the time-reversed path pairing. Notice that here we have used the properties of the trace, such that 
\begin{equation}
\nonumber
\text{tr}\left[\mathbb{R}\mathbb{R}^\dagger\right]=\frac{1}{2} \text{tr}\left[ \openone+
\left( \begin{array}{cc}
e^{i2\phi} & 0 \\
0 & e^{-i2\phi}
\end{array} \right)
\right] = 1+\cos2\phi,
\end{equation}
with 
\begin{equation}
\nonumber
\mathbb{R}=\frac{1}{2} \left[U_++U_-\right].
\end{equation}
It is worth noting that Eq.~\eqref{Gsc} overestimates the relative weight of the quantum correction. This can be seen as follows. The general expression for the semiclassical conductance of 1D loop circuits taking into account all contributing paths would read
\begin{equation}
\label{Gscfull}
G=\frac{e^2}{h}\left(g_0+\sum_{n=1}g_n\cos (2n\phi)\right).
\end{equation}

Here, $g_0$ is the (dimensionless) classical conductance with origin in path self-pairing and the sum over $n$ is the quantum correction due to time-reversed path pairing, satisfying $g_0+\sum_ng_n\le 2$ (since the conductance is upper bounded by $2e^2/h$). The comparison of the results of Fig.~\ref{fig:apC}, based on Eq.~\eqref{Gsc}, with those of Fig.~\ref{G2}(b) suggests that the first harmonic ($n=1$) dominates
over higher harmonics ($n > 1$). In particular, the numerical results of Fig.~\ref{G2}(b) can be fit accurately (up to some local extremes) by taking $g_0=0.4014$, $g_1=-0.0183$, and $g_n=0$ for $n>1$. This
fitting is obtained by minimizing the quadratic error between the semiclassical model and the numerical results. Notice that this now accounts for initial backscattering at the source contact (zero-length
paths contributing to $g_0$) that we originally neglected in the derivation of Eq.~\eqref{Gsc}. The corresponding statistical weights in Eq.~\eqref{rmn} would be $b_0=0.8889$, $b_1=b_{-1}=0.0676$ and
$b_w=0$ for $w\neq0,\pm1$, with $g_0=2-2b_0^2-4b_1b_{-1}$ and
$g_1=-4b_1b_{-1}$.

The result of Eq.~\eqref{Gsc} | corresponding to $g_0=1$, $g_1=-1$, and $g_n=0$ otherwise in~\eqref{Gscfull} | applies without significant loss of generality to any two-contact Rashba SOC circuit loop. By virtue of the analytic expressions for the AC phase $\phi$ presented in App.~\ref{AppPhases}, we have closed analytical expressions for general elliptical $N$-gons. 

As an example of our analytic findings, Fig.~\ref{fig:apC} shows the analytic conductance from Eq.~\eqref{Gsc} using the global phase given in Eq.~\eqref{eq:TotalPhase} for various polygons and for the parallel or ring. These conductance maps are presented as functions of $k_\text{R}P$ and the curvature $P/R$ (which equals $N\theta_N$ by definition). Since $G$ is proportional to $1-\cos(2\phi)$, conductance minima (maxima) occur when $\phi$ reaches even (odd) multiples of $\pi/2$. This is evident in the $k_\text{R}P$ dependence profiles of the conductance and the global phase, shown for fixed curvature values in the lower panels. The conductance maps and their associated cuts reflect the parity of the global phase $\phi(k_\text{R}P,\frac{P}{R})$ with respect to the parallel transport condition $k_\text{R}P=-\frac{P}{2R}$, depicted as a grey dashed line. Using the distance in $k_\text{R}P$ from the PT condition, $\delta_{k_\text{R}P}\equiv|k_\text{R}P+\frac{P}{2R}|$, and Eq.~\eqref{symPT0}, we observe that
\beq 
G\left(-\frac{P}{2R} +\delta_{k_\text{R}P},\frac{P}{R}\right) = G\left(-\frac{P}{2R} -\delta_{k_\text{R}P},\frac{P}{R}\right)\,.
\label{symPT}
\eeq
All cases shown in Fig.~\ref{fig:apC} share the same qualitative behavior near the PT condition. This can be understood considering first that all along the PT condition, $\delta_{k_\text{R}P}=0$, the dynamical phase of Eq.~\eqref{eq:dyn:phase} is zero, the spin texture is normal to the path and therefore the total phase becomes purely geometrical and completely determined by the solid angle of the real-space trajectory: $\phi=-\Omega_N/2$. As the curvature $P/R$ grows, the dependence of the solid angle in Eq.~\eqref{eq:solidangle} is qualitatively similar for all the closed paths shown in Fig.~\ref{fig:apC}; all these $\Omega_N$'s grow smoothly from $0$ to $2\pi$. The global phase $\phi$ starts at the planar case being $0$ producing a conductance minimum, then, as $P/R$ is increased, $\phi$ passes through $-\pi/2$ generating a conductance maximum, and finally for the maximum curvature the phase reaches $-\pi$ which again produces a conductance minimum. 
 
The second characteristic feature of the region near PT is due to the presence of the trivial identity at ${k_\text{R}P}=0$, being a robust conductance minimum for all $P/R$ (as $\phi=0$ for a spin-identity evolution $U_+=\openone$ around the loop). Note that using the symmetry relation around PT of Eq.~\eqref{symPT} the trivial identity corresponds to the $\delta_{k_\text{R}P}=\frac{P}{2R}$ case, thus implying the existence of another $\phi=0$ condition for 
 \beq
 {k_\text{R}P}=-\frac{P}{R} ~~\Rightarrow~~{k_\text{R}}=2\times\frac{-1}{2R}\,,
 \label{eq:kp:nontrivial}
 \eeq
which is twice the spin-orbit strength required for the PT condition. In Fig.~\ref{fig:apC}, the blue arrows indicate the evolution with curvature for such $\phi=0$ conditions and their associated conductance minima. It is worth noting that for nonzero curvature, such a $U_+=\openone$ condition is nontrivial as it appears at a nonzero value of ${k_\text{R}P}$. Moreover, for this non-trivial identity, both the dynamic and the geometric phases are nonzero, namely,
\bea
\phi_\text{d}=-\phi_\text{g}&=& N\theta_N \cot{\frac{\pi}{N}}\tan{\frac{\theta_N}{2}} \nonumber\\ 
&=& \frac{P}{R} \cot{\frac{\pi}{N}}\tan{\frac{P}{2NR}} \nonumber\\
&=& \frac{P}{R} \cos\vartheta_z = \frac{P}{R} \left(\Hat{B}_n\right)_z
\label{eq:phases:nti}
\eea
which follows from setting ${k_\text{R}P}$ as Eq.~\eqref{eq:kp:nontrivial} in our analytic results for the phases of Eq.~\eqref{eq:TotPhase2},~\eqref{eq:dyn:phase}, and~\eqref{eq:GeoPhase}. Note that the curvature-dependent quantity $\cos\vartheta_z$, see Eq.~\eqref{thetaZ} in App.~\ref{AppN-gons}, is the $z$ component of any of the $\hat{B}_n$ directions. This is also valid for the ring or parallel case, as for $k_\text{R}P=-\frac{P}{R}$ the global phase in Eq.~\eqref{eq:tot:parallel} is zero while the dynamic phase in Eq.~\eqref{eq:dyn:parallel} becomes $\phi_{\text{d}}^{s}=s\frac{P^2}{2\pi R^2}  = s\frac{P}{R}\cos\vartheta_z=-\phi_{\text{g}}^{s}$.

The result in Eq.~\eqref{eq:phases:nti} can be understood considering that for the $s=\uparrow$ band at ${k_\text{R}P}\rightarrow 0$ the solution is fixed in $\ketLR{z}$ all along the closed path and therefore Eq.~\eqref{eq:dynfase0} implies that 
\beq
\left.{\left\langle  \langle    h_\mathrm{poly} \rangle \right\rangle}_{\uparrow} \right|_{(k_\text{R}P)_+=0}= -\hat{z}\cdot \hat{B}_0=-\cos\vartheta_z \,.
\label{eq:averageh:ti}
\eeq
where we are using the labels defined in Eq.~\eqref{eq:symPT:labels} for the case of $\delta_{k_\text{R}P}=\frac{P}{2R}$. From the relation in  Eq.~\eqref{symPTderiv} it follows that
\beq
\left.{\left\langle  \langle    h_\mathrm{poly} \rangle \right\rangle}_{\uparrow} \right|_{(k_\text{R}P)_-=-\frac{P}{R}}= \cos\vartheta_z \,.
\label{eq:averageh:nti}
\eeq
It can be easily verified that Eq.~\eqref{eq:averageh:nti} is in full agreement with Eq.~\eqref{eq:phases:nti} by computing the dynamical phase of Eq.~\eqref{eq:dyn:averageh} for the $(k_\text{R}P)_-=-\frac{P}{R}$ solution, as this involves computing the product of  $-k_\text{R}P=+\frac{P}{R}$ with the average of Eq.~\eqref{eq:averageh:nti}.

Equations~\eqref{eq:averageh:ti} and~\eqref{eq:averageh:nti} reflect a salient relation between the spin textures of the trivial and nontrivial identity solutions. For each geodesic arc, the nontrivial identity's spin texture evolves along a parallel (with the $\hat{B}_n$ direction as its axis of symmetry) with opposite latitude sign to the trivial identity case, which maintains a fixed spin texture at $\ket{z}$. Both identity spin textures have latitudes equally separated from the "equatorial" geodesic (PT) condition at zero latitude. Notably, the nontrivial identity's spin texture rotates by a nonzero angle (of amplitude $|\frac{2}{N}(k_\text{R}P)_-|= |\frac{2P}{NR}|$) per each geodesic arc of the polygon, enabling the accumulation of a nontrivial geometric phase that precisely cancels its dynamic phase, as stated in Eq.~\eqref{eq:phases:nti}.

\end{document}